\documentclass[twocolumn]{aastex61}
\pdfoutput=1 
\usepackage{amsmath,amstext}
\usepackage[T1]{fontenc}
\usepackage{apjfonts} 
\usepackage[figure,figure*]{hypcap}
\usepackage{upgreek}
\usepackage{float}
\usepackage{rotating}
\usepackage{boldline}
\usepackage{listings}
\usepackage{hyperref}

\submitjournal{ApJ}
\received{2/1/2018}
\accepted{7/12/2018}

\shorttitle{Understanding Subdwarfs via J1256$-$0224} 
\shortauthors{Gonzales et al.}

\begin{document}

\title{Understanding Fundamental Properties and Atmospheric features of subdwarfs via a case study of SDSS J125637.13$-$022452.4 \footnote{This paper includes data gathered with the 6.5 meter Magellan Telescopes located at Las Campanas Observatory, Chile.}}

\author[0000-0003-4636-6676]{Eileen C. Gonzales}
\altaffiliation{LSSTC Data Science Fellow} 
\affiliation{Department of Astrophysics, American Museum of Natural History, New York, NY 10024, USA}
\affiliation{The Graduate Center, City University of New York, New York, NY 10016, USA}
\affiliation{Department of Physics and Astronomy, Hunter College, City University of New York, New York, NY 10065, USA}

\author[0000-0001-6251-0573]{Jacqueline K. Faherty}
\affiliation{Department of Astrophysics, American Museum of Natural History, New York, NY 10024, USA}

\author[0000-0002-2592-9612]{Jonathan Gagn\'e}
\altaffiliation{NASA Sagan Fellow}
\affiliation{Department of Terrestrial Magnetism, Carnegie Institution of Washington, Washington, DC 20015, USA}

\author{\'Etienne Artigau}
\affiliation{Institut de Recherche sur les Exoplan\`etes, D\'epartement de Physique, Universit\'e de Montr\'eal, Montr\'eal QC, H3C 3J7, Canada}

\author[0000-0001-8170-7072]{Daniella Bardalez Gagliuffi}
\altaffiliation{AMNH Kalbfleisch Fellow}
\affiliation{Department of Astrophysics, American Museum of Natural History, New York, NY 10024, USA}

\correspondingauthor{Eileen Gonzales}
\email{egonzales@amnh.org}

\begin{abstract}
We present the distance-calibrated spectral energy distribution (SED) of the sdL3.5 subdwarf SDSS J125637.13$-$022452.4 (J1256$-$0224) using its \textit{Gaia} parallax and its resultant bolometric luminosity and semi-empirical fundamental parameters, as well as updated UVW velocities. The SED of J1256$-$0224 is compared to field-age and low-gravity dwarfs of the same effective temperature ($T_\mathrm{eff}$) and bolometric luminosity.  In the former comparison, we find that the SED of J1256$-$0224 is brighter than the field source in the optical, but dims in comparison beyond the J-band, where it becomes fainter than the field from the H through W2 bands. Compared to the young source, its fainter at all wavelengths. We conclude that J1256$-$0224 is depleted of condensates compared to both objects.  A near infrared band-by-band analysis of the spectral features of J1256$-$0224 is done and is compared to the equivalent $T_\mathrm{eff}$ sample.  From this analysis, we find a peculiar behavior of the $J$-band \ion{K}{1} doublets whereby the $1.17\, \upmu$m doublet is stronger than the field or young source as expected, while the $1.25\, \upmu$m doublet shows indications of low gravity. In examining a sample of 4 other subdwarfs with comparable data we confirm this trend across different subtypes indicating that the $1.25\, \upmu$m doublet is a poor indicator of gravity for low metallicity objects.  In the $K$-band analysis of J1256$-$0224 we detect the 2.29 $\upmu$m CO line of J1256$-$0224, previously unseen in the low-resolution SpeX data. We also present fundamental parameters using \text{Gaia} parallaxes for 9 additional subdwarfs with spectral types M7 - L7 for comparison. The 10 subdwarfs are placed in a temperature sequence and we find a poor linear correlation with spectral type. We present polynomial relations for absolute magnitude in JHKW1W2, effective temperature, and bolometric luminosity versus spectral type for subdwarfs.
\end{abstract}


\section{Introduction}
Brown dwarfs are low-mass, low-temperature objects that are unable to sustain stable hydrogen burning in their cores and thus cool throughout their lifetime. With masses \mbox{$\lesssim 75~M_\mathrm{Jup}$}, brown dwarfs lie between the boundary of low mass stars and planets \citep{Saum96,Chab97}. Classified based on their red optical or near-infrared (NIR) spectra, the observed population of brown dwarfs have effective temperatures of $250-3000$ K corresponding to late-type M, L, T and Y spectral types \citep{Kirk05, Burg02a, Cush11}. 

The majority of brown dwarfs fall into three age subpopulations: field dwarfs, low surface gravity dwarfs, and subdwarfs. 
Effective temperature drives the atmospheric chemistry of brown dwarfs, thus creating differences in the observable spectroscopic and photometric features. Field dwarfs anchor the brown dwarf spectral classification methodology. Low-gravity dwarfs and subdwarfs have secondary parameters such as age, metallicity, and clouds that can alter their observable properties. Low-gravity dwarfs are identified by red near-infrared colors, weak alkali lines, and enhanced metal oxide absorption in the optical \citep{Kirk06, Kirk10, Cruz09, Alle10}. Many have been shown to be bona fide high-likelihood or candidate members of young nearby moving groups (e.g. \citealt{Fahe16, Gagn17, Liu_13, Schn16, Kell16}).  M and L subdwarfs are low-luminosity, metal-poor stars and brown dwarfs that typically display blue near-infrared $J-K_\mathrm{s}$ colors \citep{Burg03c, Burg09a}. Subdwarf spectra exhibit enhanced metal hydride absorption bands (e.g. FeH), weak or absent metal oxides (TiO, CO, VO), and enhanced collision-induced H$\,_{2}$ absorption compared to the field dwarfs (\citealt{Burg03c} and references therein). They exhibit high proper motions, substantial radial velocity, and inclined, eccentric, and sometimes retrograde Galactic orbits indicating membership in the Galactic Halo \citep{Dahn08, Burg08a, Cush09}.


In comparison to field dwarfs, the low metallicity of M and L subdwarfs leads to atmospheres containing a lower abundance of metal-oxide molecules such as TiO, VO, and CO. Consequently, this results in strong hydride bands of FeH, CaH, CrH, MgH, and AlH. The reduced abundance of these metal-oxide molecules indicates that metals in subdwarfs are prominently seen in metal-hydride species or as individual lines. With a reduced amount of metals in the atmosphere, condensate formation is difficult, thus resulting in a lack of clouds \citep{Kirk10}. This reduction in condensate formation allows metals to remain in the atmosphere to lower temperatures, thus lines of Ca\,I, \ion{Ti}{1}, \ion{Ca}{2}, and \ion{Rb}{1} are visible at temperatures lower than normally expected. An increased strength in the alkali lines of \ion{K}{1}, \ion{Fe}{1}, \ion{Rb}{1}, and \ion{Cs}{1} are also visible in subdwarf spectra. 

The metal-deficient atmosphere of subdwarfs causes increased collision-induced H$\,_{2}$ absorption in the $H$ and $K$ bands, which is believed to be what produces their blue $J-K_{s}$ color \citep{Kirk10}. This is a consequence of their higher atmospheric pressure, caused by their higher mass at a fixed temperature. Likely young objects exhibit red $J-K_\mathrm{s}$ colors corresponding to their low surface gravity \citep{Fahe16}, while subdwarfs exhibit blue $J-K_\mathrm{s}$ color corresponding to their low metallicity and likely also indicating their high surface gravity.

There are on the order of 80 subdwarfs of type sdM7 and later, which are of particular interest because they lie in the temperature range ($<3000$ K), which is comparable to directly imaged exoplanets (e.g. FW Tau b, Beta Pic b; \citealt{Krau14, Lagr10}) and they give us a handle on how low metallicity and high surface gravity differences may impact observables. For this study we use the archetypal subdwarf J1256$-$0224 as a case example for how low metallicity and high surface gravity changes observables.

Data from the literature on J1256$-$0224 is presented in Section~\ref{Pub1256}. New MKO $K$ band photometric data taken for J1256$-$0224 and new NIR FIRE spectra taken for J1256$-$0224 and 2MASS J02235464$-$5815067 (hereafter J0223$-$5815) are discussed in Section~\ref{Obs}. Section~\ref{FundParam1256} discusses how we derive the fundamental parameters of J1256$-$0224 using its distanced-calibrated SED and the \cite{Fili15} method. In Section~\ref{Comp1256}, a comparison of J1256$-$0224 with field age and low-gravity dwarfs of the same effective temperature and bolometric luminosity is completed in order to examine how the overall SED shape differs with age.  To understand how low metallicity affects spectral features, a band-by-band comparison of the spectra in the $Y$, $J$, $H$ and $K$ bands for the $T_\mathrm{eff}$ and $L_\mathrm{bol}$ samples are discussed in Section~\ref{SpecAnalysis}. Lastly, Section~\ref{Subdwarfs} places J1256$-$0224 in context with subdwarfs with parallaxes spectral typed sdM7 and later in order to better understand the subdwarf population.

\section{Published Data on J1256$-$0224}\label{Pub1256}

J1256$-$0224 was discovered by \cite{Siva09} as part of a search for L subdwarfs in the Sloan Digital Sky Survey data release 2 (SDSS DR2) spectral database. Its spectral type is discussed in \cite{Siva09}, \cite{Scholz09}, \cite{Burg09a}, \cite{Kirk16} and \cite{Zhang2017a}. \cite{Siva09}  tentatively classified J1256$-$0224 as an "sdL4". With additional data \cite{Scholz09} was able to spectral type it as $\mathrm{sdL4} \pm 1$ based on its optical spectrum, the width of its $0.77\, \upmu$m KI doublet, its blue near-infrared colors, and halo kinematics. \cite{Burg09a} revised the spectral type to sdL3.5, based on the \cite{Burg07a} method. \cite{2013MmSAI..84.1089A} classified the near-infrared spectra of J1256$-$0224 using their method as an M6 and with field-gravity, noting the stark contrast to the optical spectral type of sdL3.5. Most recently, \cite{Zhang2017a} reclassified J1256$-$0224 as an usdL3 using the $0.73-0.88\, \upmu$m region of the optical spectrum, which was based on their classification scheme anchored to the work of L\'epine et al. (2007).

Optical spectra for J1256$-$0224 are presented in \cite{Siva09} (SDSS), \cite{Scholz09} (VLT/FORS1), \cite{Burg09a} (LDSS3), \cite{Lodi15} (VLT/FORS2), and \cite{Kirk16} (Palomar/DSpec). Near-infrared spectra are presented in \cite{Burg09a} (SpeX), \cite{Martin17} (NIRSPEC) and this paper (FIRE). Some notable spectral features include: the deep \ion{K}{1} doublet at $0.77\, \upmu$m, strong bands of CrH and FeH at $0.86\, \upmu$m, atomic lines of \ion{Rb}{1}, and well defined bands of CaH and TiO near $0.705\, \upmu$m \citep{Siva09}. 

\cite{Pavlenko2016} computed synthetic SED spectra for J1256$-$0224 using a NextGen model atmosphere with $T_\mathrm{eff} = 2600$ K, $\mathrm{log}\,g = 5.0$, and $[\mathrm{Fe/H}] = -2.0$. They found that synthetic model spectra with collision-induced absorption (CIA) fit J1256$-$0224 better than models without CIA, which was most visible in the near-infrared where CIA is most significant.

\cite{Siva09} note that the $r-z$ color of J1256$-$0224 is much redder than the coolest known M subdwarfs, while the optical to NIR colors are much bluer than those of field L dwarfs. \cite{Schi09} calculated their own photometric $J$, $H$, $K_\mathrm{s}$ magnitudes.

\cite{Siva09} determined that J1256$-$0224 is a high proper motion object with $(\mu_\mathrm{\alpha},\; \mu_\mathrm{\delta}) = (-470 \pm 64,\; -378 \pm 64)$\;mas\,yr$^{-1}$. \cite{Schi09} also calculated a proper motion and determined a parallax of $11.10 \pm 2.88$ mas for J1256$-$0224. Based on its kinematics, \cite{Burg09a} designated J1256$-$0224 as a member of the Milky Way's inner halo. \cite{Burg09a} measured a radial velocity of $V_\mathrm{r} = -130\pm11$\;km\,s$^{-1}$ and determined $UVW$ velocities of $(-115\pm11,\,-101\pm18,\,-150\pm9)$\;km\,s$^{-1}$ based on a distance from the Cushing et al.\;(2009) absolute magnitude/spectral type relations. We note that the statement of improved radial velocity and space velocities of J1256$-$0224 in \cite{Lodi15} is incorrect due to their use of proper motion and parallax for SSSPM J1256$-$1408 instead of those of J1256$-$0224. With the recent release of \textit{Gaia} DR2 \citep{GaiaDR1,GaiaDR2,Lind18} the parallax and proper motions of J1256$-$0224 have improved and are listed in Table \ref{tab:1256data}.

An effective temperature value for J1256-0224 was determined in \cite{Siva09}, where they found $T_\mathrm{eff} \sim 1800$\;K based on the Burrows et al.\;(2001) evolutionary models. \cite{Schi09} determined a metallicity estimate of $[\mathrm{M/H}] = -1.3$ according to Baraffe et al.\;(1997) models. From synthetic GAIA COND-PHOENIX and DRIFT-PHOENIX model spectra and colors, \cite{Burg09a} found $T_\mathrm{eff} \sim 2100-2500$\;K, $\mathrm{log}\,g = 5.0-5.5$, and ${[\mathrm{M/H}]} = -1.5$ to $-1.0$. Using the \cite{Burg09a} spectra and NextGen atmosphere models with modified line lists, \cite{Lodi15} determined the best fit synthetic spectra to have $T_\mathrm{eff} = 2600$ K, $[\mathrm{M/H}] = -2.0$, and $\mathrm{log}\,g = 5.0$. We note that \cite{Lodi15} also determined $T_\mathrm{eff}$, $L_\mathrm{bol}$, and radius values for J1256$-$0224, however these values used the parallax and proper motion of J1256$-$1408 instead, therefore they are not valid. Lastly, \cite{Zhang2017a} derived $T_\mathrm{eff} = 2250 \pm 120$\;K, $[\mathrm{Fe/H}] = -1.8 \pm 0.2$\;dex, and $\mathrm{log}\,g = 5.50 \pm 0.2$\;dex for J1256$-$0224 from BT-Settl model fits.

\cite{Scholz09} measured the TiO5, VO-a, and PC3 spectral indices in order to classify J1256$-$0224 as an $\mathrm{sdL4} \pm 1$ using a spectral type versus spectral index relationship. \cite{Burg09a} measured the equivalent widths of H$\alpha$, \ion{Ca}{1}(6571.10 $\upmu$m), \ion{Ti}{1} (7204.47 and 8433.31$\upmu$m), \ion{Rb}{1} (7798.41 and 7945.83 $\upmu$m), \ion{Na}{1} (8182.03 and 8193.33 $\upmu$m), \ion{Cs}{1} (8519.019 and 8941.47 $\upmu$m) and \ion{Ca}{2} (8540.39 $\upmu$m) in an effort to characterize J1256$-$0224 as a subdwarf. \cite{Martin17} measured the pseudo-equivalent widths of \ion{K}{1} lines at 1.1692, 1.1778, and 1.2529 $\upmu$m and the 1.2 $\upmu$m FeH$_J$ absorption index for J1256$-$0224, which are reported in \ref{tab:1256data}. J1256$-$0224 was one of seven objects in \cite{Martin17} that spanned all of the gravity classifications in their paper. A gravity classification of INT-G (shown in Table 2 of \citealt{Martin17}) is determined based off of the median of the assigned score for the four indices measured, however they state that they exclude subdwarfs from their gravity typing. The best values for previously published data on J1256$-$0224, as well as values we present in this paper are listed Table~\ref{tab:1256data}.

\begin{deluxetable}{l c c c c}
\tablecaption{Properties of J1256$-$0224 \label{tab:1256data}}
\tablehead{\colhead{Property} &\phm{s}& \colhead{Value} &\phm{s}& \colhead{Reference}} 
  \startdata
  R.A. && $12^h 56^m 37.16^s$ && 2MASS \\ 
  Decl. && $-02 ^\circ 24' 52''.2$ && 2MASS \\
  Spectral type && sdL3.5 && 1\\
  $\pi$ (mas) && $12.55 \pm 0.72$  && 2 \\ 
  SDSS $i$ (mag) && $19.41 \pm 0.02$ && 3\\
  SDSS $z$ (mag) && $17.71 \pm 0.02$ && 3\\
  2MASS $J$ (mag) && $16.10 \pm 0.11$ && 4\\
  2MASS $H$ (mag) && $15.79 \pm 0.15$ && 4\\
  2MASS $K_{s}$ (mag) && $<15.439$ && 4\\
  $H_\mathrm{MKO}$ (mag) && $16.078 \pm 0.016$ && 5\\
  $K_\mathrm{MKO}$  (mag) && $16.605 \pm 0.099$ && 6\\
  WISE $W1$ (mag) && $15.214 \pm 0.038$ && 7\\
  WISE $W2$ (mag) && $15.106 \pm 0.098$ && 7\\
  WISE $W3$ (mag) && $<12.706$ && 7\\
  WISE $W4$ (mag) && $<8.863$ && 7\\
  {[Fe/H]} && $-1.8 \pm 0.2$ && 8 \\
  $\mu_\alpha$ (mas yr$^{-1}$) && $-511.25\pm1.33$ && 2\\  	            
  $\mu_\delta$ (mas yr$^{-1}$) && $-300.64\pm0.83$ && 2 \\ 
  $V_{r}$ (km s$^{-1}$) && $-126 \pm 10$ && 9 \\
  $V_\mathrm{tan}$ (km s$^{-1}$) && $252.7 \pm 3.9$ && 6 \\ 
  $L_\mathrm{bol}$ && $-3.63 \pm 0.05$ && 6\\
  $T_\mathrm{eff}$ (K) && $2307\pm 71$ && 6\\
  Radius ($R_\mathrm{Jup}$) && $0.94\pm0.02$ && 6\\ 
  Mass ($M_\mathrm{Jup}$ )&& $83.2\pm1.9$ && 6\\
  log $g$ (dex) && $5.37\pm0.01$ && 6\\                                                 
  Age (Gyr) && $5-10$ && 6\\
  Distance (pc) && $79.7\pm4.6$ && 6\\
  $U$ (km s$^{-1}$) \tablenotemark{a}&& $-144\pm10$ && 6\\              
  $V$ (km s$^{-1}$) \tablenotemark{a}&& $-139.1\pm8.7$ && 6\\ 
  $W$ (km s$^{-1}$) \tablenotemark{a}&& $-161.7\pm9.3 $&& 6\\ 
  EW \ion{K}{1} (1.1692 $\upmu$m) (\AA) && $7.92\pm0.34$ && 10\\
  EW \ion{K}{1} (1.1778 $\upmu$m) (\AA) && $12.3\pm0.29$ && 10\\
  EW \ion{K}{1} (1.2437 $\upmu$m) (\AA) && $2.52\pm0.78$ && 10\\
  EW \ion{K}{1} (1.2529 $\upmu$m) (\AA) && $5.06\pm0.37$ && 10\\
  FeH$_J$ (1.20 $\upmu$m) (\AA) && $1.102\pm0.01$ && 10 \\
  \enddata
\tablenotetext{a}{We present corrected UVW velocities, since \cite{Lodi15} UVW velocities used an incorrect parallax and proper motion to determine the velocities. UVW values are calculated in the local standard of rest frame in a left-handed coordinate system with U positive toward the Galactic center.}
\tablerefs{(1) \cite{Burg09a}, (2) \cite{GaiaDR1,GaiaDR2,Lind18}, (3) \cite{Adel08}, (4) \cite{Cutr03}, (5) \cite{Lawr12}, (6) This Paper, (7) \cite{Cutr12}, (8) \cite{Zhang2017a}, (9) \cite{Lodi15}, (10) \cite{Martin17}}
\end{deluxetable}

\section{Observations}\label{Obs}
\subsection{FIRE Data}
We used the 6.5m Baade Magellan telescope and the Folded-port InfraRed Echellette (FIRE; \citealt{Simcoe13}) spectrograph to obtain near-infrared spectra of J1256-0224 as well as the comparative young source J0223-5815.  Observations were made on 2016 August 12 for J1256-0224 and 2015 September 25 for J0223-5815.  For all observations, we used the echellette mode and the 0.6$\arcsec$ slit  (resolution $\lambda$/$\Delta \lambda \sim$ 6000) covering the full 0.8 - 2.5 $\micron$ band with a spatial resolution of 0.18$\arcsec$/pixel. Total exposure times of 2400s (4 ABBA nods of 600s) and 1600s (4 ABBA nods of 400s) were taken for J0223-5815 and J1256-0224 respectively.   Immediately after each science image, we obtained an A star for telluric correction and obtained a ThAr lamp spectra (HD25986 for J0223-5815 and HD110587 for J1256-0224).  At the start of the night we obtained dome flats and Xe flash lamps to construct a pixel-to-pixel response calibration.  Data were reduced using the FIREHOSE package which is based on the MASE and SpeX reduction tools (\citealt{Bochanski09}, \citealt{Cushing04}, \citealt{vacca03}). 

\subsection{Photometric Data}
MKO $K$-band observations of J1256$-$0224 were obtained at the Mont M\'egantic 1.6-m telescope with the CPAPIR infrared camera \citep{Artigau2004} on 2017 April 13.  We obtained a sequence of 100 frames, each consisting of 3 co-additions, each with a 8.1 s exposure time, for a total integration time of 40 minutes. The sequence was dithered randomly over a 3 arcminute circle. We used our custom IDL code for the data reduction. The reduction includes the division by a flat-field taken on the dome, a sky subtraction made by using a running-median of the images. The astrometric correction (second order) and registering of frames was done by cross-matching with GAIA DR1 catalog stars in the field. For the photometry, we used an aperture photometry code (a modified version of \texttt{aper.pro} in astrolib). The aperture radius is 1 FWHM and the sky radius is 2 to 4 FWHM. The FWHM and fluxes are measured directly in the final combined image. The uncertainties are estimated by taking aperture measurements in blank areas of the field and measuring the dispersion of the blank sky values. This is only correct if we are in the background-limited regime, which is the case here. The resultant photometry is listed in Table \ref{tab:1256data}. The MKO $K$ measurement provides an uncertainty and is in agreement with the previous 2MASS $K_s$ upper limit.

\section{Fundamental Parameters of J1256$-$0224}\label{FundParam1256}

\begin{figure*}
  \hspace{-0.25cm}
   \includegraphics[scale=.7]{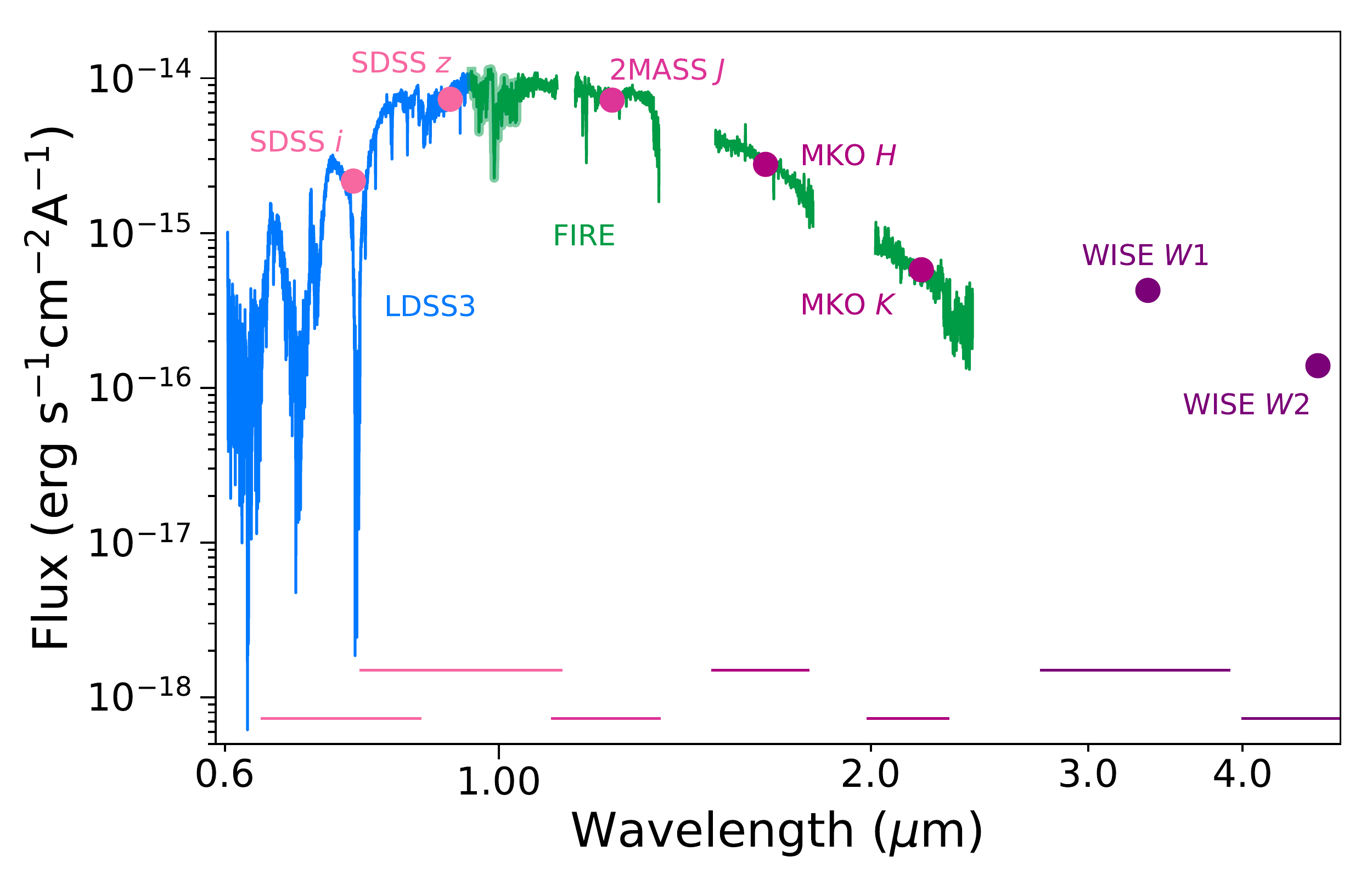}
\caption{Distance-calibrated SED of J1256$-$0224. The spectra (optical in blue, NIR in green) and photometry (shades of pink and purple) are labeled by instrument or filter system. The horizontal lines at the bottom show the wavelength coverage for the corresponding photometric measurement. The overlapping region for the optical and NIR is the fuzzy blue-green portion of the SED. Observation references can be found in Tables \ref{tab:1256data} and \ref{tab:SpectraReferences}.}
\label{fig:Regimes}
\vspace{0.5cm} 
\end{figure*}

The fundamental parameters for J1256$-$0224 were determined using the technique of \cite{Fili15}, where we create a distance-calibrated spectral energy distribution (SED) using the spectra, photometry, and parallax\footnote{SEDkit is available on GitHub at \url{https://github.com/hover2pi/SEDkit}}. The SED of J1256$-$0224 uses the LDSS3 optical spectrum from \cite{Burg09a}, near-infrared FIRE spectrum from this paper, and the photometry (excluding 2MASS $K_s$) and parallax listed in Table \ref{tab:1256data}.

Using the optical and NIR spectra, a composite spectrum is constructed and scaled to the absolute magnitudes in each filter. The overlapping red optical region in the range $0.95-1.03\, \upmu$m was combined as an average. Details for SED generation can be found in \cite{Fili15}. The SED of J1256$-$0224 is shown in Figure \ref{fig:Regimes}, with the various components labeled. The bolometric luminosity is determined by integrating under the SED. The effective temperature is calculated using a radius determined from the Saumon \& Marley 2008 low metallicity (-0.3 dex) cloudless evolutionary model along with the bolometric luminosity using the Stefan-Boltzmann law. An age range of $5-10$ Gyr was used for J1256$-$0224, which conservatively encompasses possible subdwarf ages. 

Using the semi-empirical approach, we derived the following parameters: $L_\mathrm{bol} = -3.63\pm 0.05$, $T_\mathrm{eff} = 2307\pm71$\;K, $R = 0.94\pm0.02$\;$R_\mathrm{Jup}$, $M = 83.2\pm1.9 \;M_\mathrm{Jup}$, $\mathrm{log}\,g = 5.37\pm0.01$\;dex. If the Saumon \& Marley 2008 solar metallicity cloudless models were instead used, the derived effective temperature would be 36~K cooler. The fundamental parameters derived for J1256$-$0224 can be seen in Table \ref{tab:1256data}. Our values for the bolometric luminosity and radius differ from those of \cite{Lodi15}, because they have used an incorrect parallax value in their calculations. Our derived values for $T_\mathrm{eff}$ and $\mathrm{log}\,g$ are consistent with the model based values from \cite{Zhang2017a}.

From their metallicity versus effective temperature relation, \cite{Zhang2017b} classified J1256$-$0224 as a brown dwarf i.e. a mass below the hydrogen burning boundary. Using their relation, we too would classify J1256$-$0224 as a brown dwarf. We concluded this using our $T_\mathrm{eff}$ and mass values, along with a $[\mathrm{Fe/H}] = -1.8$ for J1256$-$0224 and the hydrogen burning minimum mass for $[\mathrm{Fe/H}] = -1.8$ from \cite{Zhang2017a}. We note that on the metallicity vs effective temperature figure from \cite{Zhang2017b}, field objects of the same effective temperature as J1256$-$0224 are late-M dwarfs. We also found late-M type field dwarfs were comparison sources for the same $T_\mathrm{eff}$ as J1256$-$0224.

\section{A Comparative Sample for J1256$-$0224}\label{Comp1256}

\subsection{Sample Selection and properties}
\begin{deluxetable*}{l c c c c c c c c} 
\tablecaption{The $T_\mathrm{eff}$ and $L_\mathrm{bol}$ Samples\label{tab:TeffLbolsample}}
\tablehead{\colhead{R.A.} & \colhead{Decl.} & \colhead{Designation} &\phm{s}& \colhead{Shortname} & \colhead{Discovery Ref.} & \colhead{Opt. SpT} & \colhead{Spt Ref.}} 
  \startdata
  12 56 37.13& $-$02 24 52.4 & SDSS J125637.13$-$022452.4 && J1256$-$0224& 1 & sdL3.5 & 2\\ \hline
  &&&$T_\mathrm{eff}$ Sample&&&&\vspace{0.1cm} \\ \hline
  00 24 24.63& $-$01 58 20.1 & 2MASS J00242463$-$0158201 && J0024$-$0158 & 3 & M9.5 & 4 \\             
  20 00 48.41& $-$75 23 07.0 & 2MASS J20004841$-$7523070 && J2000$-$7523 & 5 & M9\,$\gamma$ & 6\\ \hline      
  &&&$L_\mathrm{bol}$ Sample&&&&\vspace{0.1cm} \\ \hline
  10 48 14.64& $-$39 56 06.2 & DENIS--P J1048.0$-$3956 && J1048$-$3956 & 7 & M9 & 7 \\             
  02 23 54.64& $-$58 15 06.7 & 2MASS J02235464$-$5815067 && J0223$-$5815 & 8 & L0\,$\gamma$ & 9\\              
  \enddata
\tablerefs{(1) \cite{Siva09}, (2) \cite{Burg09a}, (3) \cite{Cruz02}, (4) \cite{Kirk95}, (5) \cite{Cost06}, (6) \cite{Fahe16}, (7) \cite{Mont01}, (8) \cite{Reid08b}, (9) \cite{Cruz09}}
\end{deluxetable*}

\begin{deluxetable*}{l c c c c c c c c}
  \tablecaption{Properties of the $T_\mathrm{eff}$ and $L_\mathrm{bol}$ samples \label{tab:PropertiesTeffLbol}}
  \tablehead{\colhead{Property} & \colhead{J0024$-$0158} & \colhead{Ref.} & \colhead{J2000$-$7523} & \colhead{Ref.} &     \colhead{J1048$-$3956} & \colhead{Ref.} & \colhead{J0223$-$5815} & \colhead{Ref.}}
  \startdata
  $\pi$ (mas) & $86.6 \pm 4.0$ & 1 & $33.95 \pm 0.15$ & 2 & $247.22 \pm 0.12$ & 2 & $24.42 \pm 0.60$ & 2 \\
  SDSS $g$ & $21.07 \pm 0.04$ & 3 & $\cdots$ & $\cdots$ & $\cdots$ & $\cdots$ & $\cdots$ & $\cdots$ \\
  SDSS $r$ & $18.68 \pm 0.01$ & 3 & $\cdots$ & $\cdots$ & $\cdots$ & $\cdots$ & $\cdots$ & $\cdots$ \\
  SDSS $i$ & $16.18 \pm 0.004$ & 3 & $\cdots$ & $\cdots$ & $\cdots$ & $\cdots$ & $\cdots$ & $\cdots$ \\
  SDSS $z$ & $14.37 \pm 0.004$ & 3 & $\cdots$ & $\cdots$ & $\cdots$ & $\cdots$ & $\cdots$ & $\cdots$ \\
  $V$ & $20.01 \pm 0.051$ & 4 & $21.157 \pm 0.008$ & 5 & $17.532 \pm 0.057$ & 6 & $\cdots$ & $\cdots$ \\
  $R$ & $17.45 \pm 0.016$ & 4 & $18.379 \pm 0.001$ & 5 & $15.051 \pm 0.014$ & 6 & $\cdots$ & $\cdots$ \\
  $I$ & $15.0 \pm 0.025$ & 4 & $16.119 \pm 0.024$ & 5 & $\cdots$ & $\cdots$ & $\cdots$ & $\cdots$ \\
  2MASS $J$ & $11.992 \pm 0.035$ & 7 & $12.734 \pm 0.026$ & 7 & $9.538 \pm 0.022$ & 7 & $15.07 \pm 0.048$ & 7\\
  2MASS $H$ & $11.084 \pm 0.022$ & 7 & $11.967 \pm 0.027$ & 7 & $8.905 \pm 0.044$ & 7 & $14.003 \pm 0.036$ & 7\\
  2MASS $K_{s}$ & $10.539 \pm 0.023$ & 7 & $11.511 \pm 0.026$ & 7 & $8.447 \pm 0.023$ & 7 & $13.42 \pm 0.042$ & 7\\
  MKO $J$ & $11.73 \pm 0.03$ & 8 & $\cdots$ & $\cdots$ & $\cdots$ & $\cdots$ & $\cdots$ & $\cdots$ \\
  MKO $H$ & $11.1 \pm 0.03$ & 8 & $\cdots$ & $\cdots$ & $\cdots$ & $\cdots$ & $\cdots$ & $\cdots$ \\
  MKO $K$ & $10.53 \pm 0.03$ & 8 & $\cdots$ & $\cdots$ & $\cdots$ & $\cdots$ & $\cdots$ & $\cdots$ \\
  MKO $L^\prime$ &$9.78 \pm 0.13$ & 9 & $\cdots$ & $\cdots$ & $\cdots$ & $\cdots$ & $\cdots$ & $\cdots$ \\
  WISE $W1$ & $10.166 \pm 0.024$ & 10 & $11.108 \pm 0.023$ & 10 & $8.103 \pm 0.024$ & 10 & $12.819 \pm 0.024$ & 10 \\
  WISE $W2$ & $9.9 \pm 0.019$ & 10 & $10.797 \pm 0.02$ & 10 & $7.814 \pm 0.021$ & 10 & $12.431 \pm 0.024$  & 10\\
  WISE $W3$ & $9.412 \pm 0.039$ & 10 & $10.55 \pm 0.069$ & 10 & $7.462 \pm 0.018$ & 10 & $11.64 \pm 0.15$ & 10\\
  WISE $W4$ & $\cdots$ & $\cdots$ & $\cdots$ & $\cdots$ & $7.226 \pm 0.087$ & 10 & $\cdots$ & $\cdots$ \\
  IRAC [3.6 $\upmu$m] & $9.94 \pm 0.03$ & 11 & $\cdots$ & $\cdots$ & $\cdots$ & $\cdots$ & $\cdots$ & $\cdots$ \\
  IRAC [4.5 $\upmu$m] & $9.91 \pm 0.03$ & 11 & $\cdots$ & $\cdots$ & $\cdots$ & $\cdots$ & $\cdots$ & $\cdots$ \\
  IRAC [5.8 $\upmu$m] & $9.72 \pm 0.01$ & 11 & $\cdots$ & $\cdots$ & $\cdots$ & $\cdots$ & $\cdots$ & $\cdots$ \\
  IRAC [8.0 $\upmu$m] & $9.55 \pm 0.01$ & 11 & $\cdots$ & $\cdots$ & $\cdots$ & $\cdots$ & $\cdots$ & $\cdots$ \\
  $L_\mathrm{bol}$ & $-3.44 \pm 0.04$ & 12& $-3.04 \pm 0.006$ & 12 & $-3.489 \pm 0.002$ & 12 & $-3.615 \pm 0.082$ & 12 \\
  $T_\mathrm{eff}$ (K) & $2385 \pm 77$ & 12 & $2388 \pm 36$ & 12 & $2328 \pm 54$ & 12 & $1819 \pm 90$ & 12 \\
  Radius ($R_\mathrm{Jup}$) & $1.09 \pm 0.05$ & 12 & $1.72 \pm 0.05$ & 12 & $1.08 \pm 0.05$ & 12 & $1.53 \pm 0.04$ & 12 \\ 
  Mass ($M_\mathrm{Jup}$) & $79 \pm 11$ & 12 & $29.6 \pm 3.0$ & 12 & $78 \pm 10$ & 12 & $19.1 \pm 2.2$ & 12 \\
  log $g$ (dex) & $5.21 \pm 0.10$ & 12 & $4.37 \pm 0.07$ & 12 & $5.21 \pm 0.10$ & 12 & $4.27 \pm 0.04$ & 12 \\              
  Age (Gyr)\tablenotemark{a} & $500-10000$ & 12 & $21-27$ & 12 & $500-10000$ & 12 & $41-49$ & 12 \\
  Distance (pc) & $11.55 \pm 0.53$ & 12 & $29.45 \pm 0.13$ & 12 & $4.03 \pm 0.01$ & 12 & $36.5 \pm 3.5$& 12 \\
  \enddata
  \tablecomments{Photometric points with uncertainties greater than 0.15 magnitudes were excluded from the construction of the SED. The effective temperature is determined based off age estimates and evolutionary models.}
  \tablenotetext{a}{We assume ages for field dwarfs to be in the range $500-10000$ Myr and ages for nearby young moving groups from \cite{Bell15}.}
    \tablerefs{(1) \cite{Tinn95}, (2) \cite{GaiaDR1,GaiaDR2,Lind18}, (3) \cite{Ahn_12}, (4) \cite{Diet14}, (5) \cite{Cost06}, (6) \cite{Cost05}, (7) \cite{Cutr03}, (8) \cite{Legg02a}, (9) \cite{Legg98}, (10) \cite{Cutr12}, (11) \cite{Patt06}, (12) This Paper}
\end{deluxetable*}

\begin{deluxetable*}{l c c c c c c c c c}
\tablecaption{Spectra used to construct SEDs of the $T_\mathrm{eff}$ and $L_\mathrm{bol}$ samples\label{tab:SpectraReferences}}
\tablehead{\colhead{Name} & \colhead{OPT} & \colhead{OPT} & \colhead{OPT} & \colhead{NIR} & \colhead{NIR} & \colhead{ NIR} & \colhead{MIR} & \colhead{MIR} & \colhead{MIR} \vspace{-.1cm}\\
 & & \colhead{Obs. Date} & \colhead{Ref.} & & \colhead{Obs. Date} & \colhead{Ref.} & &\colhead{Obs. Date} & \colhead{Ref.}}   
  \startdata
  J1256$-$0224 & LDSS3 & 2006--05--07 & 1 & FIRE & 2016--08--13 & 2 & $\cdots$ & $\cdots$ & $\cdots$\\
  J0024$-$0158 & KPNO 4m: MARS & 2003--07--10 & 3 & SpeX SXD & 2009--12--01 & 4,5 & IRS & 2004--01--04 & 6\\
  J2000$-$7523 & CTIO 4m: R--C Spec & 2003--04--23 & 7 & FIRE & 2013--07--28 & 8 & $\cdots$ & $\cdots$ & $\cdots$\\
  J1048$-$3956 & CTIO 1.5m: R--C Spec& 2003--05--15 & 7 & SpeX SXD & 2009--12--01 & 4 & $\cdots$ & $\cdots$ & $\cdots$\\
  J0223$-$5815 & GMOS--S & 2005--08--19 & 7 & FIRE & 2015--12--21 & 2 & $\cdots$ & $\cdots$ & $\cdots$\\
  \enddata
\tablerefs{(1) \cite{Burg09a}, (2) This Paper, (3) \cite{Cruz07}, (4) \cite{Rayn09}, (5) \cite{Cush05}, (6) \cite{Cush06b}, (7) \cite{Reid08b}, (8) \cite{Fahe16}}
\end{deluxetable*}

In order to better understand the fundamental parameters of J1256$-$0224 and how they affect its SED, two comparative samples were constructed. The first sample consists of objects with similar effective temperature and the second consists of objects with similar bolometric luminosity. We did not compare objects of the same spectral type because, within the same subclass, secondary parameters (e.g. metallicity) influence the spectral features. To try to disentangle these secondary parameters, we chose to look at objects with similar fundamental parameters. Effective temperature was chosen as a comparison parameter because it strongly affects the atmospheric chemistry. Bolometric luminosity was chosen to examine how the same amount of flux is distributed differently across wavelengths among objects of various radii and hence various ages. When these two parameters are fixed, clouds and metallicity remain the likely causes for differences between objects in the samples. For both comparative samples, we chose a low-gravity (likely young) object and a field-age object.

Individual sources were chosen from the samples of \cite{Fahe16}, which looked at a large sample of low-gravity dwarfs, and \cite{Fili15}, which examined both field and low gravity objects. The bolometric luminosities in both the \cite{Fahe16} and \cite{Fili15} samples were empirically derived, while their effective temperatures are semi-empirical derived using radii from the Saumon \& Marley 2008 hybrid cloud models. In addition, objects from both samples were required to have medium-resolution NIR data ($\lambda/\Delta\lambda > 1000$ at J band), in order to resolve features in a band-by-band comparative analysis of the spectra. Field dwarf comparison SEDs chosen from \cite{Fili15} were reconstructed with the same data used in that work, with the exception of replacing their low-resolution SpeX data with medium-resolution NIR data in this work and using updated \textit{Gaia} DR2 parallaxes when available. Low-gravity dwarf comparison SEDs from \cite{Fahe16} were constructed with new NIR FIRE data or re-reduced spectra as well as updated \textit{Gaia} DR2 parallaxes. The low-gravity source in the $T_\mathrm{ef}$ sample and both comparison objects in the $L_\mathrm{bol}$ sample originally included synthetic photometric values in their SEDs, where details on how these values were calculated are discussed in \cite{Fili15}. Only measured photometric values were used in the generation of the SEDs in our samples. These three modifications result in differences up to 1$\sigma$ in the values reported by \cite{Fili15} and \cite{Fahe16}.

Our comparative sample of similar effective temperatures consists of 2MASS J00242463$-$0158201 (hereafter J0024$-$0158) and 2MASS J20004841$-$7523070 (hereafter J2000$-$7523). J0024$-$0158 is a field brown dwarf discovered by \cite{Cruz02} and spectral typed as an M9.5 in both the optical by \cite{Kirk95} and the infrared by \cite{Geba02}. J2000$-$7523 was discovered by \cite{Cost06} and is stated to be an M9\,$\gamma$ in both the optical and IR in \cite{Fahe16}. As reported in \cite{Fahe16}, J2000$-$7523 is a member of the $\beta$ Pictoris moving group and thus has an age ranging from $21-27$~Myr \citep{Bell15}. 

Our comparative sample of similar bolometric luminosity sources consists of DENIS--P J1048.0$-$3956 (hereafter J1048$-$3956) and 2MASS J02235464$-$5815067 (hereafter J0223$-$5815). J1048$-$3956 was discovered and optically spectral typed as an M9 by \cite{Mont01}. J1048$-$3956 was discovered by \cite{Reid08b} and optically spectral typed as a L0\,$\gamma$ by \cite{Cruz09}. As reported in \cite{Fahe16}, J0223$-$5815 is a member of the Tucana-Horlogium moving group with an age ranging from $41-49$~Myr \citep{Bell15}. Discovery and spectral type references for both samples are listed in Table~\ref{tab:TeffLbolsample}. Properties, such as photometry and parallax, for the $T_\mathrm{eff}$ and $L_\mathrm{bol}$ samples are listed in Table~\ref{tab:PropertiesTeffLbol}, while spectra used in the SED construction is listed in Table~\ref{tab:SpectraReferences}.

\begin{figure*}
\centering
 \includegraphics[scale=.7]{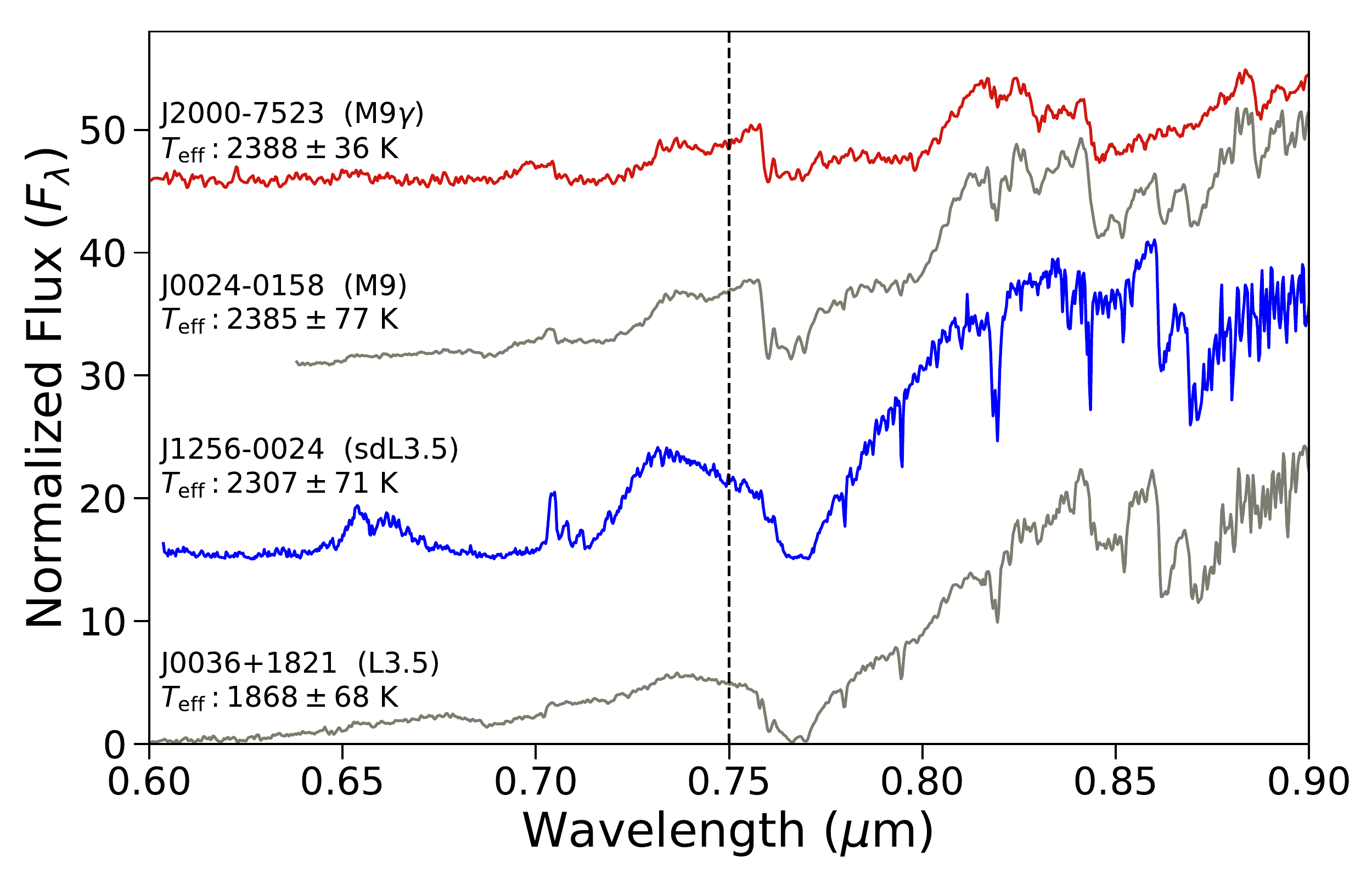}
\caption{Comparison of J1256$-$0224 to objects of the same effective temperature and a field object of the same spectral Type. All spectra were resampled to the same dispersion relation using a wavelength-dependent Gaussian convolution and are normalized based on the average flux between 0.64 $\upmu$m and 0.65 $\upmu$m and offset by a constant (15,30,45). In the red optical, J1256$-$0224 resembles the M9 and M9 $\gamma$ prior to 0.75 $\upmu$m, while it is visibly more similar to the L3.5 beyond 0.75 $\upmu$m.}
\label{fig:redoptical}
\end{figure*}

\section{Spectral Analysis}\label{SpecAnalysis}
Using their own spectral-typing system, \cite{Zhang2017a} found that L subdwarfs have temperatures between 100 and 400 K higher compared to field L dwarfs of the same spectral type, dependent on the metallicity subclass and spectral subtype of the subdwarf. Subdwarfs can therefore have effective temperatures similar to objects classified 2--3 subtypes earlier. \cite{Zhang2017a} noted that in addition to metallicity differences, there is a large spread in mass, $\mathrm{log}\,g$, and age with any spectral type for subdwarfs. NIR spectra are mainly affected by $T_\mathrm{eff}$ and metallicity, while optical spectra are most sensitive to $T_\mathrm{eff}$ \citep{Zhang2017a}. 

As stated in \cite{Siva09}, J1256$-$0224 appears to display a combination of late-M and mid-L spectral features, which is most prominently seen in the $0.6 - 0.9\, \upmu$m region. Figure \ref{fig:redoptical} contains a field- and low-gravity source of the same effective temperature, and a field dwarf of the same spectral type as J1256$-$0224 for comparison. At $0.75\, \upmu$m, there is a spectral type transition where at shorter wavelengths, the spectrum of J1256$-$0224 appears more similar to the M9 and longwards of $0.75\, \upmu$m the spectra matches better to the L3.5 dwarf. Due to this combination of M and L dwarf spectral features, we tested for the presence of a binary companion using the SpeX prism spectrum for J1256$-$0224 and the technique of \cite{Bard14}. The results showed a 28\% confidence that the binary hypothesis is statistically significant over the single hypothesis, i.e J1256$-$0224 is not a spectral binary.

In this section, the SED of J1256$-$0224 is placed into context by comparing its overall shape to young- and field-age sources with equivalent temperatures and bolometric luminosities. We also utilized the optical and NIR medium resolution spectral data to compare the red optical and individual $Y$, $J$, $H$, and $K$ bandpasses of each source, similar to the analysis of \cite{Fahe14}.

\subsection{Comparison of full SEDs for the fixed-$T_\mathrm{eff}$ sample}

\begin{figure*}
\centering
 \includegraphics[scale=.7]{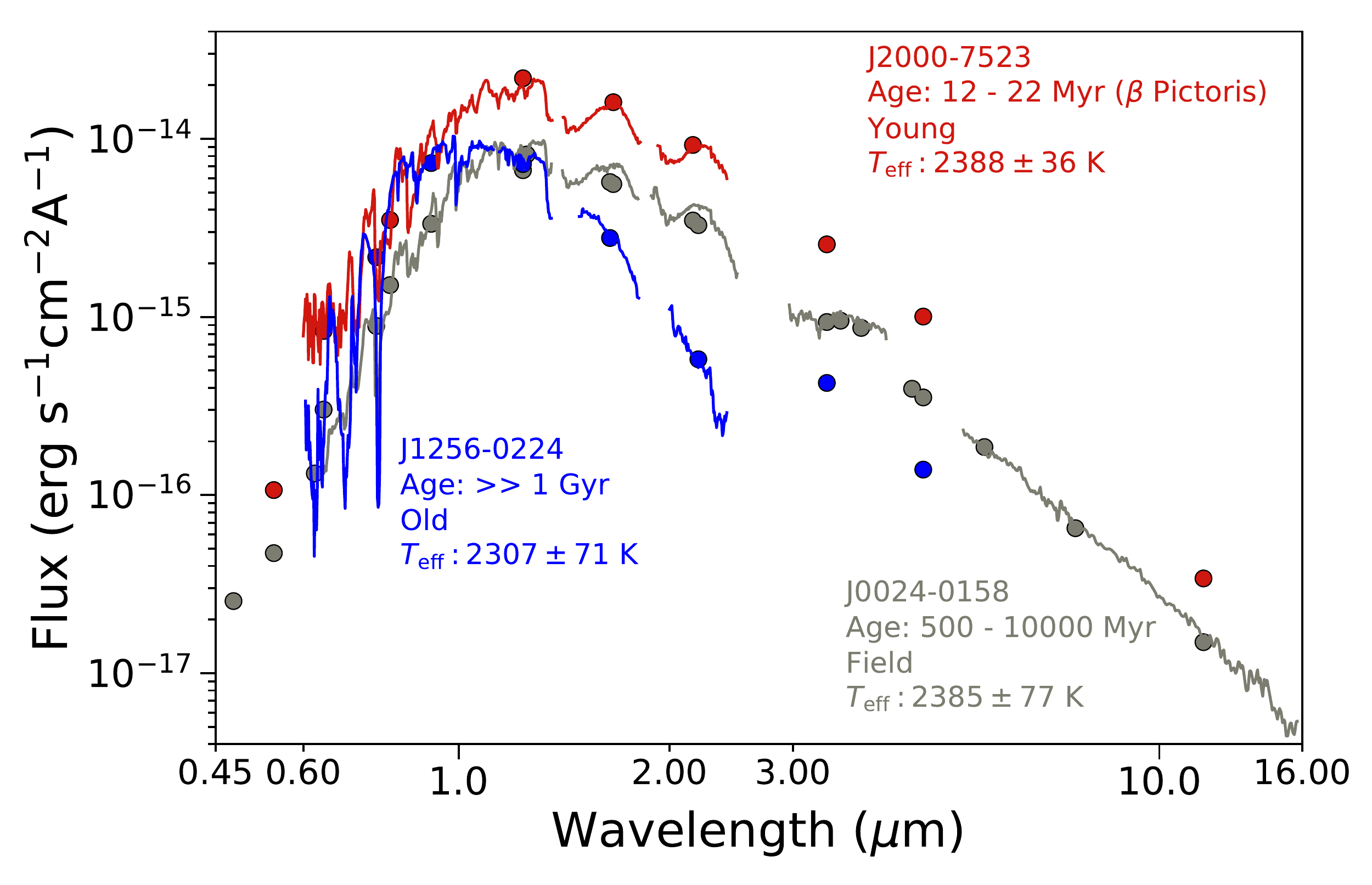}
\caption{Distance-calibrated SEDs of a J2000$-$7523 (low-gravity, red), J0024$-$0158 (field-age, grey), and J1256$-$0224 (subdwarf, blue). Objects are approximately the same effective temperature. All spectra were resampled to the same dispersion relation using a wavelength-dependent Gaussian convolution.}
\label{fig:compTeff}
\vspace{0.5cm} 
\end{figure*}

Figure \ref{fig:compTeff} contains the full SEDs of the fixed-$T_\mathrm{eff}$ sample across the $0.45-16\, \upmu$m region. J1256$-$0224 is brighter than the field source in the optical, but dims in comparison beyond the $J$-band, where it becomes fainter than the field from mid $J$ through $W2$ bands. While there are vague similarities around $1\, \upmu$m between J1256$-$0224 and J2000$-$7523, the later is far brighter at nearly all wavelengths. Previous work found evidence that subdwarfs are likely cloudless, while young sources have thicker or higher-lying clouds \citep{Fahe12, Fahe16}. Differing atmospheric conditions between these three equivalent temperature sources shown in Figure \ref{fig:compTeff} may be representative of how the atmosphere of J1256$-$0224 is depleted of condensates, while J2000$-$7526 is enhanced compared to the field dwarf. \cite{Fahe12} compared Saumon \& Marley 2008 evolutionary models, with a cloud sedimentation parameter and three gravities to the NIR color-magnitude diagram for L and T dwarfs. They found by varying the sedimentation parameter, that the models could fit portions of L and T dwarfs with various degrees. \cite{Fahe12} showed that the cloudless model was the only one that reached the location of the two L subdwarfs (J0532+8246 and J1626+3925 in the NIR color-magnitude diagram. \cite{Fahe12} also examined the effect of metallicity by using the Burrows 2006 evolutionary models, with cloudless and cloudy models of varying metallicites. The low-metallicity, cloudless models were the only ones to cross through the space of the subdwarfs. They also showed that for the Burrows models, the largest factor in fitting the L and T dwarfs was the clouds, with metallicity having a smaller effect. \cite{Zhang2017a} fit J1256$-$0224 with the BT-Settl 2014 model grid over a temperature range of $1400-2600$ K, $-2.5\leq [\mathrm{Fe/H}]\leq -0.5$, and $5.0\leq$log~$g\leq5.75$. The best fit for J1256$-$0224 was found to have a metallicty of $-1.8$ and a log $g$ of 5.5. Both \cite{Fahe12} and \cite{Zhang2017a} try to address the atmospheric features with the current models, however there are currently no models that can allow us to vary gravity, metallicty, and clouds at the same time. With atmospheric retrieval codes such as \textit{Brewster} \citep{Burn17}, we can begin to test for these parameters simultaneously on J1256$-$0224 (Gonzales et al. in prep). 

\subsection{Band-by-Band Analysis: Fixed-$T_\mathrm{eff}$ sample}
\begin{figure}
\hspace{-0.3cm}
 \includegraphics[scale=0.36]{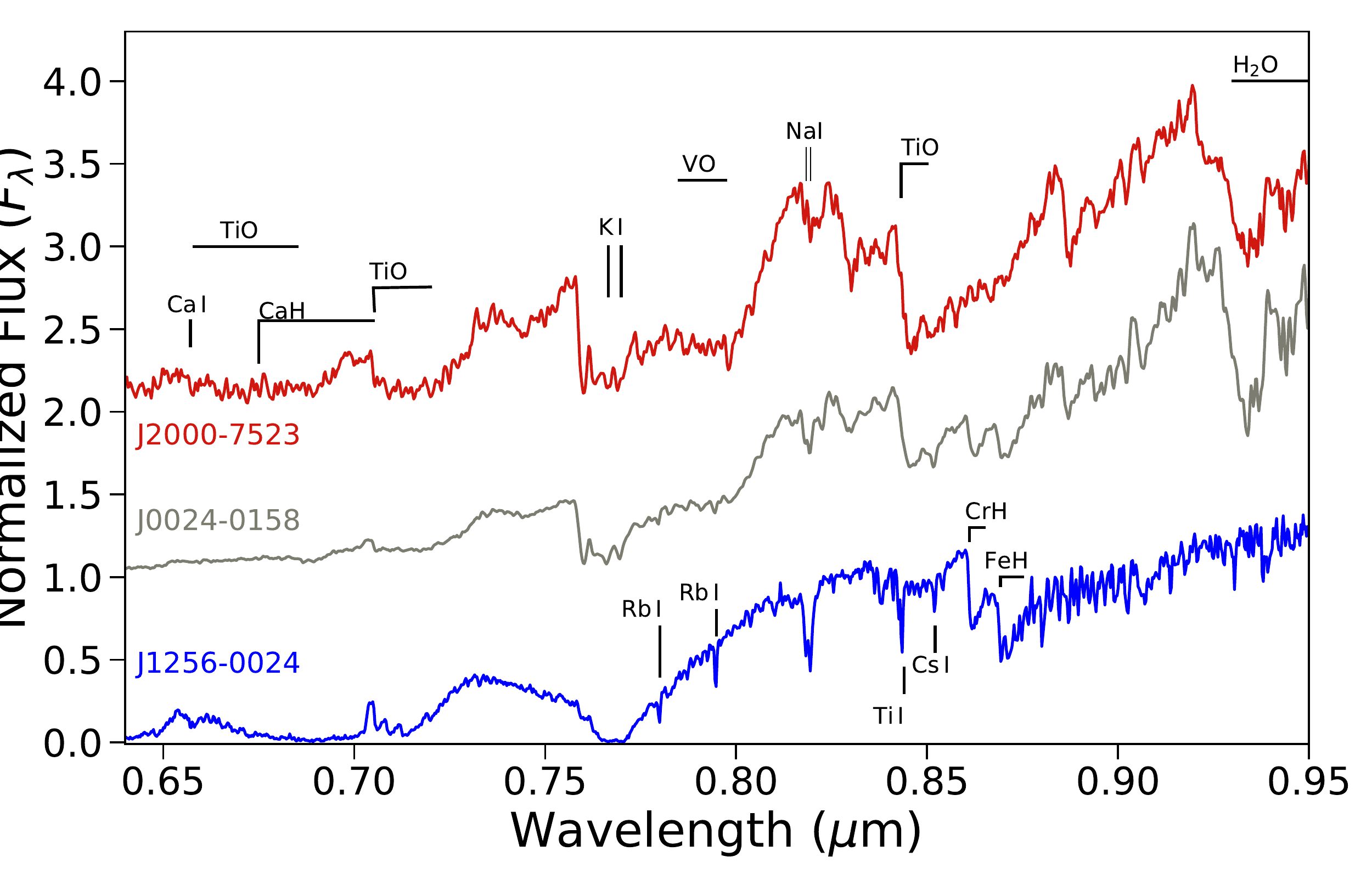}
\caption{Optical comparison of effective temperature sample. J2000$-$7523 (low-gravity, red), J0024$-$0158 (field-age, grey), and J1256$-$0224 (subdwarf, blue). All spectra were resampled to the same dispersion relation using a wavelength-dependent Gaussian convolution and are offset by a constant. Spectra were normalized by the average over $0.825-0.840\, \upmu$m.}
\label{fig:Opticalbandbyband}
\vspace{0.5cm} 
\end{figure}

\subsection{Optical}
Figure \ref{fig:Opticalbandbyband} shows the $0.64-0.95\, \upmu$m optical data with various features labeled. The \ion{Ca}{1} feature near $0.66\, \upmu$m is visible in the spectrum of J1256$-$0224, while it is not clearly seen in either the field- or low-gravity sources. The TiO band near $0.7\,\upmu$m is affects all three sources, for J1256$-$0224 the band is not as wide as the J0024$-$0158 and has a steeper slope near $0.73\,\upmu$m. J2000$-$7523 has a broader TiO feature than J1256$-$0224 and has an almost vertical shape near $0.73\,\upmu$m. The plateau between $\sim0.73-0.76\,\upmu$m for J1256$-$0224 has a blue slope, while there is a visible dip in the center for both the field- and low-gravity sources where it is broadest for the field dwarf. The \ion{K}{1} doublet is broad for all sources, however the VO for both the low-gravity and field objects affects the longward end of the broadened doublet. When comparing the \ion{Na}{1} doublets, it is shallowest for J0024$-$0158, slightly deeper for J200$-$7523, and deepest for J1256$-$0224. There is no indication of TiO absorption near $0.85\,\upmu$m or H$_2$O starting near $0.93\,\upmu$m for J1256$-$0224, while both features are seen in the field- and low-gravity sources. 

\begin{figure*}
\gridline{\fig{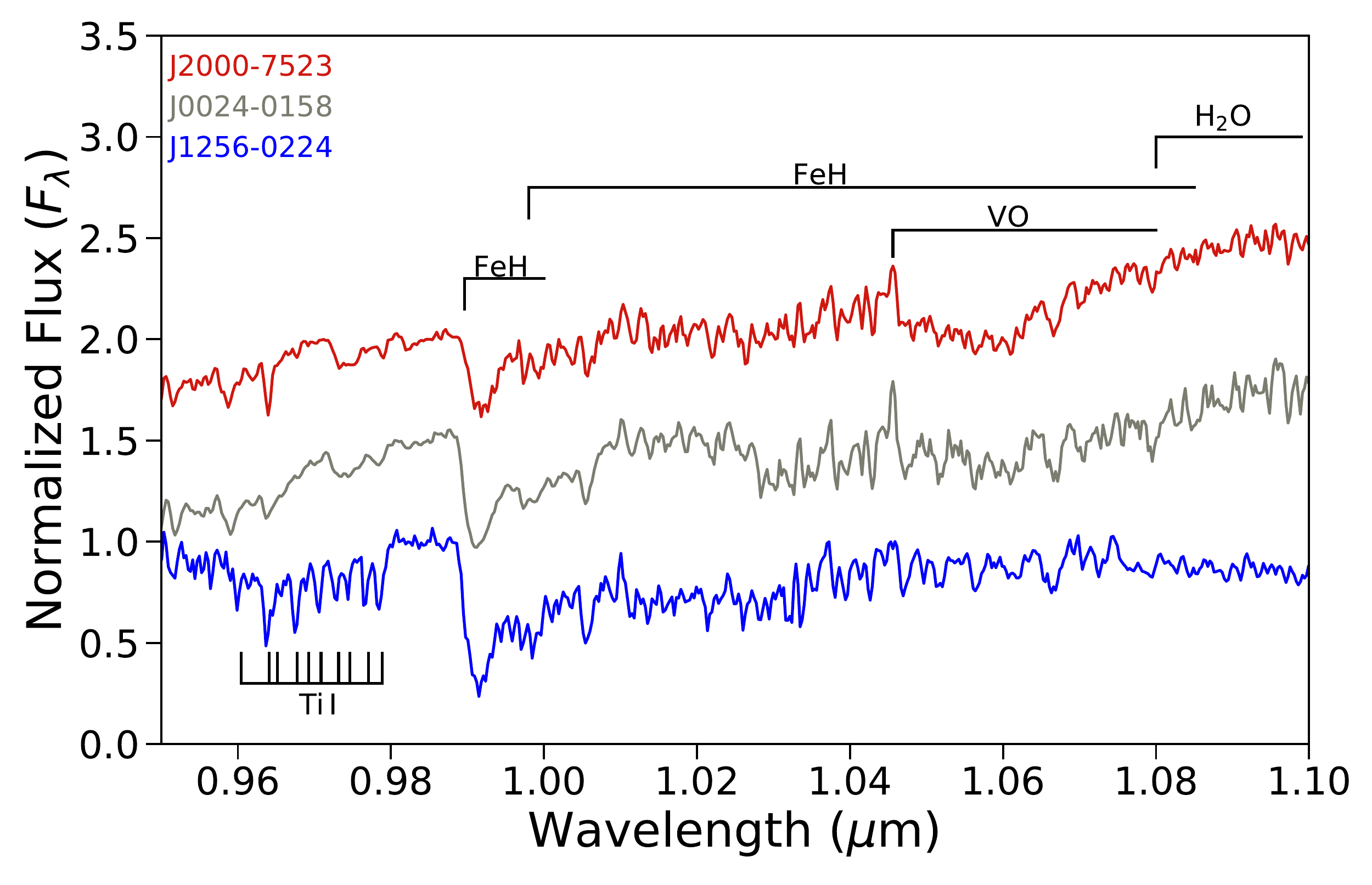}{0.5\textwidth}{\large(a)}
          \fig{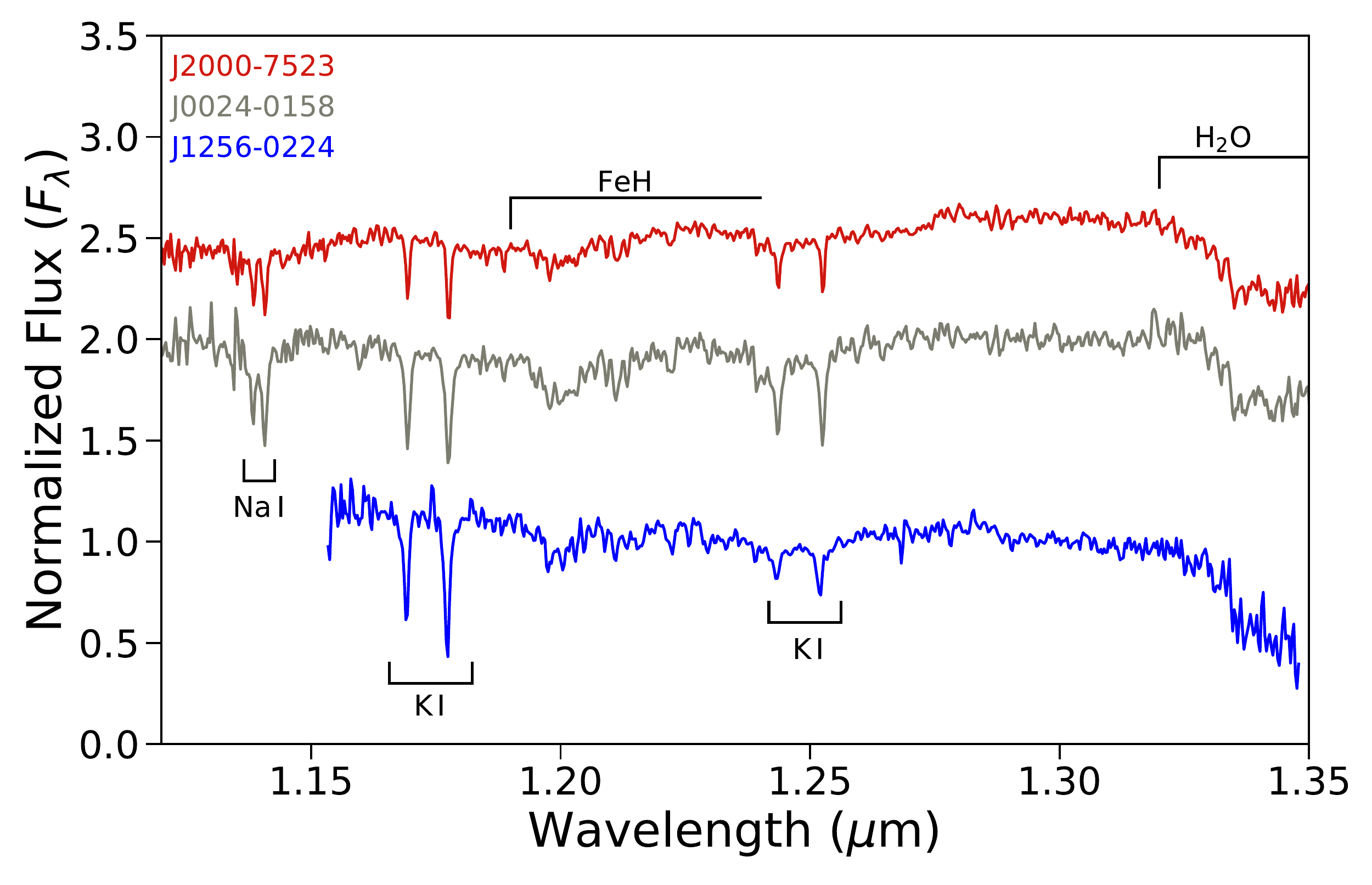}{0.5\textwidth}{\large(b)}} 
\gridline{\fig{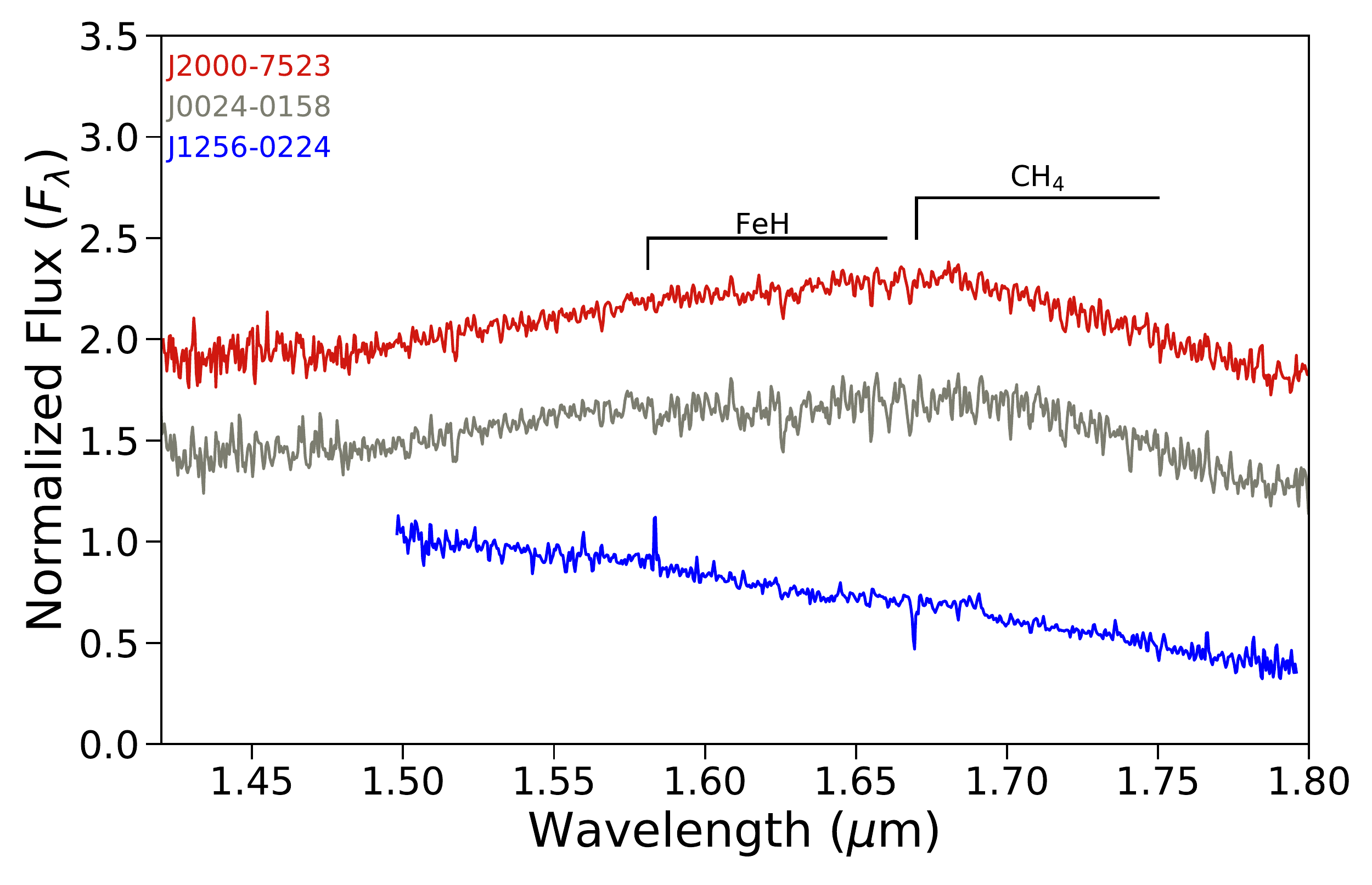}{0.5\textwidth}{\large(c)}
          \fig{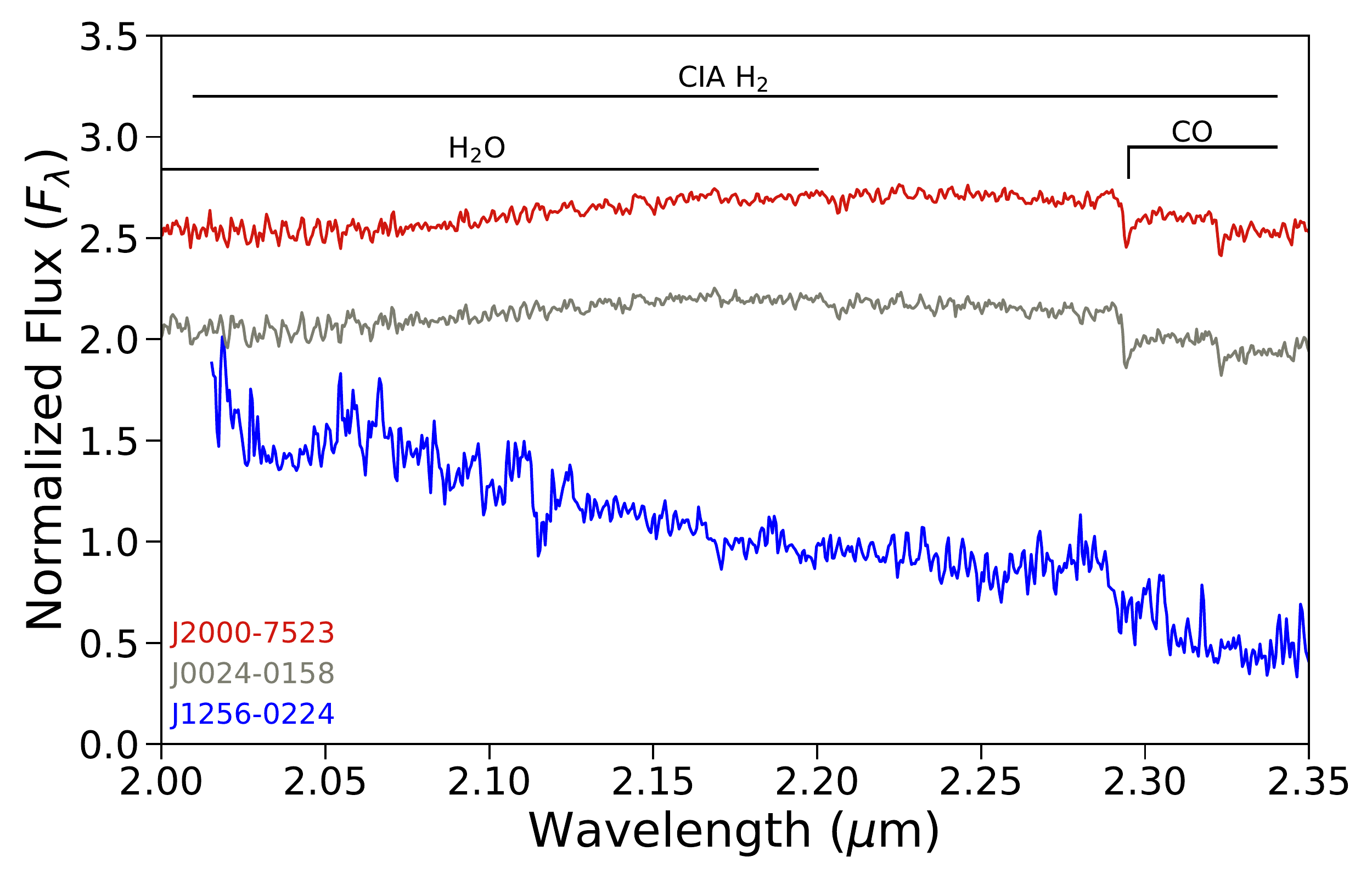}{0.5\textwidth}{\large(d)}}
\caption{Band-by-band comparison of effective temperature sample. J2000$-$7523 (low-gravity, red), J0024$-$0158 (field-age, grey), and J1256$-$0224 (subdwarf, blue). All spectra were resampled to the same dispersion relation using a wavelength-dependent Gaussian convolution and are offset by a constant. (a) $Y$ band. Spectra are normalized by the average flux taken across $0.98-0.988\, \upmu$m. (b) $J$ band. Spectra are normalized by the average flux taken across $1.29-1.31\, \upmu$m. (c) $H$ band. Spectra are normalized by the average flux taken across $1.5-1.52\, \upmu$m. (d) $K$ band. Spectra are normalized by the average flux taken across $2.16-2.20\, \upmu$m.}
\label{fig:Teffbandbyband}
\vspace{0.5cm} 
\end{figure*}

\subsubsection{$Y$ band}
Figure \ref{fig:Teffbandbyband}a shows the $0.95-1.10\, \upmu$m  $Y$-band data with \ion{Ti}{1}, FeH, VO, and H$_2$O features labeled. The spectra for the $Y$ band were normalized by averaging the flux taken across the $0.98-0.988\, \upmu$m region, which is relatively featureless.  A noticeable difference between these three objects is the \ion{Ti}{1} lines in the range $0.96-0.98\, \upmu$m, where many lines are visible in the spectrum of J1256$-$0224, while only two of the \ion{Ti}{1} lines are easily visible for J2000$-$7523 and J0024$-$0158. These lines are are indicative of inhibited condensate formation \citep{Burg03c, Rein06}. Another noticeable feature is the depth of the Wing-Ford FeH band at $0.9896\, \upmu$m, which increases in depth with the age of the object, with J1256$-$0224 having the deepest band. The Wing-Ford band is very strong in M dwarfs, however here we note that it is stronger in J1256$-$0224 indicating that J1256$-$0224 is likely cloudless. With the lack of clouds in J1256$-$0224, we are able to probe a deeper layer of the atmosphere at the Wing-Ford FeH band, as would be seen for T dwarfs. T dwarfs are thought to be cloudless due to dust settling below the photosphere at $\sim 1000$ K \citep{Burg02a}. J1256$-$0224 also lacks the VO absorption from $1.05-1.08\, \upmu$m and the redward slope caused by the $1.08-1.10\, \upmu$m H$_2$O band as seen in both J2000$-$7523 and J0024$-$0158.

\subsubsection{$J$ band}
Figure \ref{fig:Teffbandbyband}b shows the $1.12-1.35\, \upmu$m $J$-band data with FeH and H$_2$O molecular features, as well as the alkali doublets of \ion{K}{1} and \ion{Na}{1} labeled. The $J$-band spectra were normalized by the average flux over the featureless $1.29-1.31\, \upmu$m region. The $1.12-1.15\, \upmu$m portion of the spectrum of J1256$-$0224 is not shown due to the large amount of noise due to telluric absorption in this region, which obscures the \ion{Na}{1} line. Therefore, we cannot compare this particular \ion{Na}{1} line to the sample objects. We can however compare the \ion{K}{1} doublets. The most noticeable aspect of these lines is the distinct difference in the depth of the \ion{K}{1} doublet near $1.17\, \upmu$m and the depth of the doublet near $1.25\, \upmu$m. The deep blue doublet indicates that J1256$-$0224 is a high-gravity object, while the shallow red doublet indicates low gravity. \cite{Martin17} used the equivalent widths of both \ion{K}{1} doublets along with the FeH$_J$ index and also found that the equivalent width for the red \ion{K}{1} doublet indicates a low gravity. Therefore, the \ion{K}{1} doublet near $1.17\, \upmu$m is not a good gravity indicator for J1256$-$0224 which warrants further investigation. An equivalent width measurement of the \ion{K}{1} doublets for this spectrum can be found in Alam et al.\,(in prep). To see if the \ion{K}{1} doublet near $1.17\, \upmu$m behaves this way for subdwarfs in general, we examine subdwarfs in our subdwarf sample that have medium-resolution $J$-band data in Section \ref{SubdwarfKIlines}. We also note that the H$_2$O band of J1256$-$0224 is slightly deeper than both the field and low gravity comparisons.

\begin{figure*}
\centering
 \includegraphics[scale=.7]{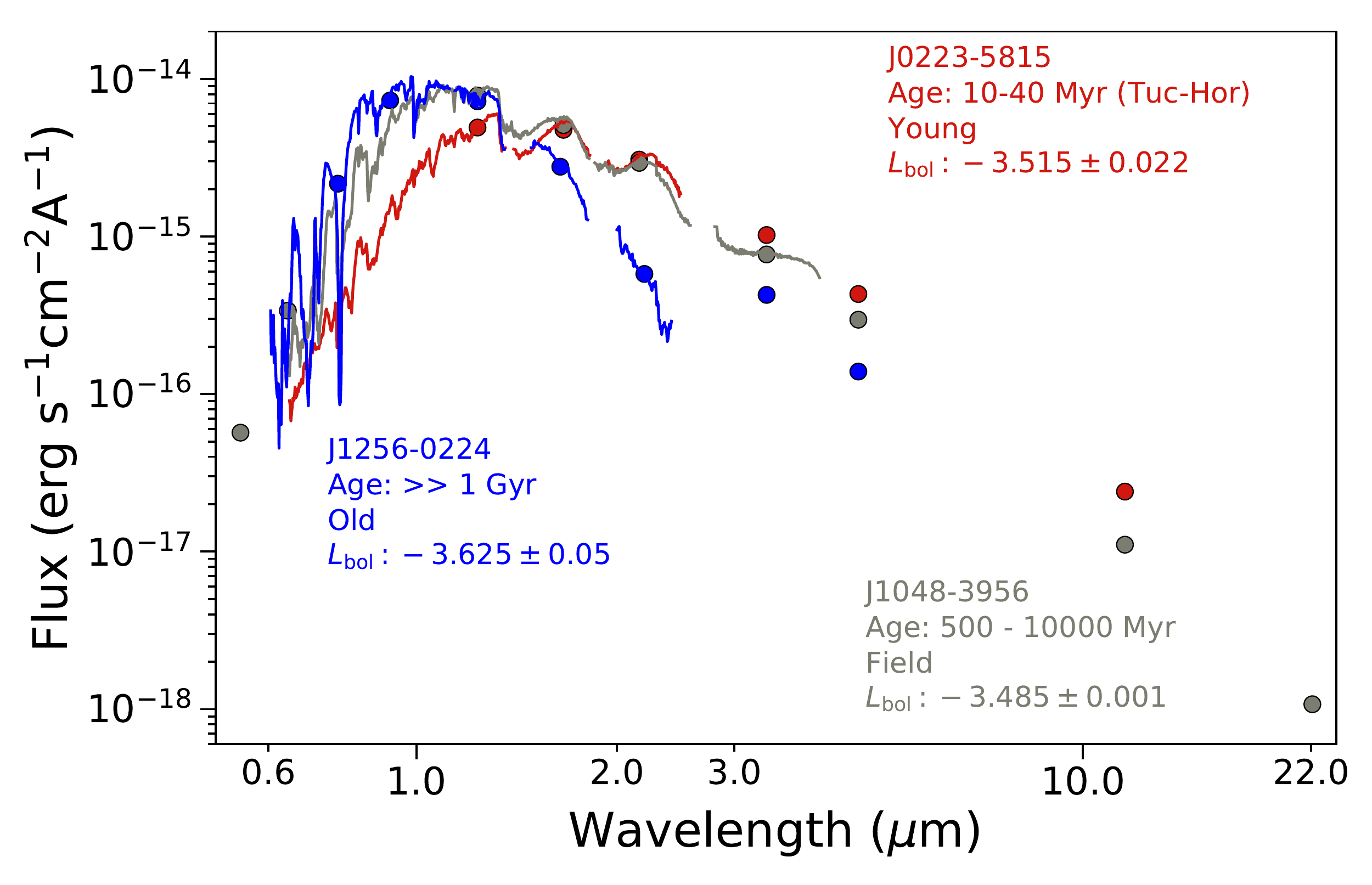}
\caption{Distance calibrated SEDs of J0223$-$5815 (low-gravity, red), J1048$-$3956 (field-age, grey), and J1256$-$0224 (subdwarf, blue). All have similar (within $< 0.15$ dex) bolometric luminosity. All spectra were resampled to the same dispersion relation using a wavelength-dependent Gaussian convolution.}
\label{fig:compLbol}
\vspace{0.5cm} 
\end{figure*}

\subsubsection{$H$ band}\label{Hbandteff}
Figure \ref{fig:Teffbandbyband}c shows the $1.42-1.80\, \upmu$m $H$-band data with FeH, and CH$_4$ molecular features labeled. The $H$-band spectra were normalized by the average flux over the featureless $1.5-1.52\, \upmu$m region. The $H$-band shape of J1256$-$0224 is significantly different from those of J2000$-$7523 and J0024$-$0158. Gravity and $K$-band collision-induced H$_2$ absorption shapes the longer wavelength end of the $H$-band. However, this difference in shape could also be due to the varying metallicity of these objects. \cite{Zhang2017a} shows that the $H$- and $K$-band flux is significantly decreased for L subdwarfs compared to L dwarfs. Figure 9 of \cite{Zhang2017a} shows this best, with the decrease in flux seen as you go through their metallicity subtypes. Since J1256$-$0224 is typed as an usdL3 in the \cite{Zhang2017a} system, you would expect this dramatic decrease in H band flux due to the extremely low metallicity.

\subsubsection{$K$ band}
Figure \ref{fig:Teffbandbyband}d shows the $2.0-2.35\, \upmu$m $K$-band data with H$_2$O, CO, and collision-induced H$_2$ absorption features labeled. The $K$-band spectra were normalized by the average flux over the $2.16-2.20\, \upmu$m region due to the relatively flat spectral region. The slope from $2.051-2.283\,\upmu$m for J1256$-$0224 may be due to systematics from the reduction with this order. The $K$ band slope for our FIRE data does not match the SpeX spectra from \cite{Burg09a}, however the rest of our NIR spectrum slope matches, thus we caution the $K$- band slope in this region. The most notable feature in the $K$ band is the $2.3\, \upmu$m CO absorption. Previously, \cite{Burg09a} examined a low-resolution SpeX prism spectrum of J1256$-$0224 and found no evidence for CO, where weakened or absent CO is a tell tale sign for subdwarfs \citep{Burg03c}. The CO absorption is comparable between the young object J2000$-$7523 and the field object J0024$-$0158. J1256$-$0224 displays clear CO absorption, but it is weakened in comparison.

\subsection{Comparing objects of the same bolometric luminosity}
Figure \ref{fig:compLbol} shows the full SEDs of the fixed-$L_\mathrm{bol}$ sample across the $0.5-22\, \upmu$m region. When comparing objects of the same bolometric luminosity, differences in their radii are most apparent. From the optical to $Y$ band, a spread in the flux is displayed, where J1256$-$0224 is over-luminous compared to J1048$-$3956, and J0223$-$5815 is under-luminous compared to J1048$-$3956. The flux of J1256$-$0224 and J1048$-$3956 overlap in the $J$ band. From the $H$ band out to the mid-infrared (MIR), J1256$-$0224 becomes under-luminous compared to the field object J1048$-$3956. The young low-gravity object J0223$-$5815 is over-luminous compared to J1048$-$3956 in the $K$ band through the MIR. Such a redistribution of flux to the optical for J1256$-$0224 points to further evidence for a lack of clouds.

A band-by-band analysis with the $L_\mathrm{bol}$ sample was not informative as spectral features behave similarly to the $T_\mathrm{eff}$ sample, thus the corresponding figures are not reported here.

\section{J1256$-$0224 in context with parallax-calibrated subdwarfs}\label{Subdwarfs}

\begin{deluxetable*}{l c c c c c c c c c c c c c c} 
\tablecaption{The Subdwarf Sample\label{tab:Subdwarfsample}}
\tablehead{\colhead{R.A.} & \colhead{Decl.} & \colhead{Designation} & \colhead{Shortname} & \colhead{Discovery Ref.} & \colhead{SpT} & \colhead{SpT Ref.} & \colhead{$\pi$ (mas)} & \colhead{Pi Ref.}} 
  \startdata
  12 56 37.13 & $-$02 24 52.4 & SDSS J125637.13$-$022452.4 & J1256$-$0224& 1 & sdL3.5 & 2 & $12.55 \pm 0.72$ & 3\\ \hline 
  05 32 53.46 & $+$82 46 46.5 & 2MASS J05325346+8246465 & J0532+8246 & 4 & sdL7 & 2 & $40.24 \pm 0.64$ & 3\\            
  06 16 40.06 & $-$64 07 19.5 & 2MASS J06164006$-$6407194 & J0616$-$6407& 5 & sdL5 & 5 & $19.85 \pm 6.45$ & 6 \\          
  10 13 07.35 & $-$13 56 20.4 & SSSPM J1013$-$1356 & J1013$-$1356 & 7 & sdM9.5 & 7 & $17.87 \pm 0.22$ & 3\\ 
  12 56 14.06 & $-$14 08 39.6 & SSSPM J1256$-$1408 & J1256$-$1408 & 1\tablenotemark{a} & sdM8 & 1\tablenotemark{a} & $22.52 \pm 0.20$ & 3\\ 
  14 25 05.11 & $+$71 02 09.7 & LSR J1425+7102 & J1425+7102 & 8 & sdM8 & 8 & $13.63 \pm 0.13$ & 3\\ 
  14 39 00.31 & $+$18 39 38.5 & LHS 377 & $\cdots$ & 9 & sdM7 & 10 & $25.75 \pm 0.10$ & 3\\ 
  14 44 20.67 & $-$20 19 22.3 & SSSPM J1444$-$2019 & J1444$-$2019 & 11 & sdM9 & 11 & $57.80 \pm 0.56$ & 12\\ 
  16 10 29.00 & $-$00 40 53.0 & LSR J1610$-$0040 & J1610$-$0040 & 13 & sdM7 & 2 & $30.73 \pm 0.34$ & 14\\ 
  16 26 20.34 & $+$39 25 19.1 & 2MASS J16262034+3925190 & J1626+3925 & 1 & sdL4 & 2 & $32.49 \pm 0.23$ & 3\\ 
  20 36 21.65 & $+$51 00 05.2 & LSR J2036+5059 & J2036+5059 & 15 & sdM7.5 & 16 & $23.10 \pm 0.29$ & 12\\ 
  \enddata
\tablenotetext{a}{Discovered and spectral typed by Scholz et al. via \cite{Schi09}.}
\tablerefs{(1) \cite{Schi09}, (2) \cite{Burg07a}, (3) \cite{GaiaDR1,GaiaDR2,Lind18}, (4) \cite{Burg03c}, (5) \cite{Cush09}, (6) \cite{Fahe12}, (7) \cite{Scho04a}, (8) \cite{Lepi03b}, (9) \cite{Lieb79}, (10) \cite{Gizi97}, (11) \cite{Scho04c}, (12) \cite{Dahn17}, (13) \cite{Lepi03c}, (14) \cite{Kore16}, (15) \cite{Lepi02}, (16) \cite{Lepi03a}}
\end{deluxetable*}

There are 11 subdwarfs of spectral type sdM7 or later that have the necessary observational data to generate distance-calibrated SEDs. Here we present these SEDs to put J1256$-$0224 in context with the other known subdwarfs with parallaxes. Our sample includes 7 subdwarfs typed M7--M9.5 and 4 L subdwarfs ranging from L3.5--L7, including J1256$-$0224. Four subdwarfs in our sample have optical, NIR, and MIR spectra, four have optical and NIR spectra, one has optical and MIR spectra, while the remaining subdwarfs have an optical or NIR spectra only.  Most of the sample had 2MASS and WISE photometry while some objects also had SDSS or IRAC photometry. 

The \cite{Kirk16} optical spectrum for J1256$-$1408 is not publicly available and therefore we could not complete its SED. The subdwarf sample SEDs were generated using the Saumon \& Marley 2008 low metallicity (-0.3 dex) cloudless models. If instead the Saumon \& Marley 2008 solar metallicity cloudless models were used the derived temperatures would be on average $\sim25$~K cooler. Table~\ref{tab:Subdwarfsample} contains the full subdwarf sample and their parallaxes, Tables~\ref{tab:subdwarfphot}--\ref{tab:subdwarfphotIRAC} contain the photometry, and Table~\ref{tab:subdwarfspectra} contain spectra used for generating the SEDs.

\begin{deluxetable*}{l c c c c c c c}
\tablecaption{2MASS and WISE Photometry for subdwarf comparison SEDs \label{tab:subdwarfphot}}
\tablehead{\colhead{Name} & \colhead{$J$} & \colhead{$H$} & \colhead{$K_\mathrm{s}$} & \colhead{$W1$} & \colhead{$W2$} & \colhead{$W3$} & \colhead{References}} 
  \startdata
  J0532+8246 & $15.179 \pm 0.058$ & $14.904 \pm 0.091$ & $14.92 \pm 0.15$ & $\cdots$ & $\cdots$ & $\cdots$ & 1 \\ 
  J0616$-$6407 & $16.40 \pm 0.11$ & $16.28 \pm 0.23$ & $\cdots$ & $ 15.65 \pm 0.03$ & $15.183 \pm 0.042$  & $\cdots$ & 1, 2\\             
  J1013$-$1356 & $14.621 \pm 0.032$ & $14.382 \pm 0.049$ & $14.398 \pm 0.078$ & $13.782 \pm 0.028$ & $13.545 \pm 0.035$ & $\cdots$ & 1, 3\\
  J1256$-$1408 & $14.011 \pm 0.027 $ & $13.618 \pm 0.033$ & $13.444 \pm 0.037$ & $13.118 \pm 0.026$ & $12.896 \pm 0.028$ & $\cdots$& 1, 3\\          
  J1425+7102 & $14.775 \pm 0.037$ & $14.405 \pm 0.054$ & $14.328 \pm 0.092$ & $13.892 \pm 0.025$ & $13.66 \pm 0.03$ & $\cdots$ & 1, 3\\
  LHS 377 & $13.194 \pm 0.029$ & $12.73 \pm 0.03$ & $12.479 \pm 0.025$ & $12.298 \pm 0.027$ & $12.051 \pm 0.025$ & $11.67 \pm 0.11$ & 1, 3\\
  J1444$-$2019 & $12.546 \pm 0.026$ & $12.142 \pm 0.026$ & $11.933 \pm 0.026$ & $11.464 \pm 0.024$ & $11.211 \pm 0.022$ & $10.967 \pm 0.09$ & 1,4\\
  J1610$-$0040 & $12.911 \pm 0.022$ & $12.302 \pm 0.022$ & $12.302 \pm 0.022$ & $11.639 \pm 0.025$ & $11.639 \pm 0.025$ & $11.639 \pm 0.025$ & 1, 3\\
  J1626+3925 & $14.435 \pm 0.029$ & $14.533 \pm 0.050$ & $14.466 \pm 0.074$ & $13.461 \pm 0.025$ & $13.091 \pm 0.028$ & $\cdots$ & 1, 3\\
  J2036+5059 & $13.611 \pm 0.029$ & $13.160 \pm 0.036$ & $12.936 \pm 0.033$ & $\cdots$ & $\cdots$ & $\cdots$ & 1\\
  \enddata
\tablecomments{References listed as: 2MASS, WISE. Photometry with uncertainties greater than 0.25 mags were not included in the SED.}
\tablerefs{(1) \cite{Cutr03}, (2) \cite{Kirk14}, (3) \cite{Cutr12}, (4) \cite{Cutr13}}
\end{deluxetable*}

\begin{deluxetable*}{l c c c c c c c c c c c c}
\tablecaption{SDSS Photometry for subdwarf comparison SEDs \label{tab:subdwarfphotSDSS}}
\tablehead{\colhead{Name} & \phm{s} & \colhead{$u$} & \phm{s} &\colhead{$g$} & \phm{s} & \colhead{$r$} & \phm{s} & \colhead{$i$} & \phm{s} & \colhead{$z$}& \phm{s} & \colhead{Reference}} 
  \startdata
  J0532+8246 && $\cdots$ && $\cdots$ && $22.43 \pm 0.26$ && $19.966 \pm 0.054$ && $17.04 \pm 0.02$ && 1\\
  LHS 377 && $21.89 \pm 0.13$ && $19.384 \pm 0.011$ && $17.681 \pm 0.006$ && $15.696 \pm 0.005$ && $14.707 \pm 0.006$ && 2 \\
  J1610$-$0040 && $\cdots$ && $\cdots$ && $17.976 \pm 0.007$ && $15.903 \pm 0.004$ && $14.663 \pm 0.005$ && 3\\
  J1626+3925 && $\cdots$ && $23.08 \pm 0.16$ && $20.65 \pm 0.04$ && $17.92 \pm 0.01$ && $16.13 \pm 0.01$ && 4\\ 
  \enddata 
  \tablecomments{Photometry with uncertainties greater than 0.3 mags were not included in the SED.}
\tablerefs{ (1) \cite{Alam15}, (2) \cite{Ahn_12}, (3) \cite{Adel09}, (4) \cite{Adel08}}
\end{deluxetable*}

\begin{deluxetable*}{l c c c c c c c c c c}
\tablecaption{IRAC Photometry for subdwarf comparison SEDs \label{tab:subdwarfphotIRAC}}
\tablehead{\colhead{Name} &\phm{strings} & \colhead{[3.6]} & \phm{strings} & \colhead{[4.5]} & \phm{strings} & \colhead{[5.8]} & \phm{strings} & \colhead{[8.0]}  & \phm{strings} & \colhead{Reference}} 
  \startdata
  J0532+8246 && $13.37 \pm 0.03$ && $13.22 \pm 0.02$ && $13.23 \pm 0.1 $ && $13.03 \pm 0.1$ && 1 \\ 
  J1626+3925 && $\cdots$ && $\cdots$ && $\cdots$ && $12.82 \pm 0.18$ && 2\\ 
  \enddata
\tablerefs{(1) \cite{Patt06}, (2) Spitzer PID251}
\end{deluxetable*}

\begin{deluxetable*}{l c c c c c c c c c}
\tablecaption{Spectra used to construct the subdwarf sample SEDs \label{tab:subdwarfspectra}}
\tablehead{\colhead{Name} & \colhead{OPT} & \colhead{OPT} & \colhead{OPT } & \colhead{NIR} & \colhead{NIR} & \colhead{NIR} & \colhead{MIR} & \colhead{MIR} & \colhead{MIR} \vspace{-.1cm}\\
 & &\colhead{Obs. Date} &\colhead{Ref.} & & \colhead{Obs. Date} & \colhead{Ref.}  &\colhead{Obs. Date} & \colhead{Ref.}}   
  \startdata
  J0532+8246 & LRIS & 2003--01--03 & 1 & NIRSPEC & 2002--12--24 & 1 & IRS & 2005--03--23& 2\\
  J0616$-$6407 & GMOS-S & 2007--09--13 & 3 & X-Shooter & 2016--01--24& 4 & $\cdots$ & $\cdots$ & $\cdots$\\
  J1013$-$1356 & GMOS-N & 2004--11--21 & 5 & SpeX & 2004--03--12& 6 & IRS & 2005--06--07& 7\\
  J1256$-$1408 &$\cdots$ & $\cdots$ & $\cdots$ & $\cdots$ & $\cdots$ & $\cdots$ & $\cdots$ & $\cdots$ & $\cdots$\\
  J1425+7102 & KPNO 4m: R--C Spec & 20002--05--19 & 8 & $\cdots$ & $\cdots$ & $\cdots$ & IRS & 2005--06--04 & 7\\
  LHS 377 & X-Shooter (UVB,VIS) & 2014--02--20 & 9 & X-Shooter & 2014--02--20 & 9 & IRS & 2005--07--01 & 7 \\
  J1444$-$2019 & EFOSC2 & 2004--03--16 & 10 & SpeX & 2005--03--23 & 11 & $\cdots$ & $\cdots$ & $\cdots$\\
  J1610-0040 & MkIII & 2003--02--19 & 12 & SpeX & 2003--07--06 & 13 & $\cdots$ & $\cdots$ & $\cdots$\\
  J1626+3925 & OSIRIS & $\cdots$\tablenotemark{a} & 14 & SpeX & 2004--07--23 & 6 & IRS & 2005--06--03 & 7\\
  J2036+5059 & KAST &2001-12-09 & 15  & SpeX & 2003--10--06 & 13 & $\cdots$ & $\cdots$ & $\cdots$ \\
  \enddata
\tablenotetext{a}{Combined spectrum of observations on 2011--02--05, 2011--02--12, and 2011--04--13.}
\tablerefs{(1) \cite{Burg03c}, (2) Spitzer PID51, (3) \cite{Cush09}, (4) \cite{Zhang2017a}, (5) \cite{Burg07a}, (6) \cite{Burg04c}, (7) Spitzer PID251, (8) \cite{Lepi03b}, (9) \cite{Rajp16}, (10) \cite{Scho04c}, (11)\cite{Bard14}, (12) \cite{Lepi03c}, (13) \cite{Cush06a}, (14) \cite{Lodi15}, (15) \cite{Lepi03a}}
\end{deluxetable*}

\begin{deluxetable*}{l c c c c c c c c}
\tablecaption{Subdwarf sample fundamental parameters determined with SEDkit \label{tab:subdwarfSEDout}}
\tablehead{\colhead{Name} & \colhead{Optical SpT} & \colhead{L$_\mathrm{bol}$} & \colhead{T$_\mathrm{eff}$} & \colhead{Radius} & \colhead{Mass} & \colhead{log(g)} &\colhead{Age} & \colhead{Distance} \vspace{-.1cm}\\ 
 & & & \colhead{(K)} & \colhead{(R$_\mathrm{J}$)} & \colhead{(M$_\mathrm{J}$)} & \colhead{(dex)} & \colhead{(Gyr)} & \colhead{(pc)} }
  \startdata
  J1256$-$0224 & sdL3.5 & $-3.625\pm 0.05$ & $2307\pm71$ & $0.94\pm0.02$ & $83.2\pm1.9$ & $5.37\pm0.01$ & $5-10$ & $79.7\pm4.6$  \\   \hline
  J0532+8246 & sdL7 & $-4.277 \pm 0.015$ & $1677 \pm 25$ & $0.84 \pm 0.02$ & $72.1 \pm 5.2 $ & $5.4 \pm 0.05$ & $5-10$ & $24.85 \pm 0.39$\\
  J0616$-$6407 & sdL5 & $-4.24 \pm 0.28$ & $1720 \pm 280$ & $0.84 \pm 0.04$ & $70.5 \pm 9.2$ & $5.39 \pm 0.08$ & $5-10$ & $50 \pm 16$\\
  J1013$-$1356 & sdM9.5 & $-3.307 \pm 0.012$ & $2621 \pm 22$ & $1.05 \pm 0.01$ & $91.89 \pm 0.55$ & $5.32 \pm 0.05$ & $5-10$ & $55.96 \pm 0.69 $\\
  J1256$-$1408 & sdM8 & $\cdots$ & $\cdots$ & $\cdots$ & $\cdots$ & $\cdots$ & $5-10$ & $44.4 \pm 0.39$ \\ 
  J1425+7102 & sdM8 & $-3.059 \pm 0.011$ & $2829 \pm 22$ & $1.2 \pm 0.01$ & $104.48 \pm 0.76$ & $5.25 \pm 0.05$ &  $5-10$ & $73.36 \pm 0.71 $ \\
  LHS 377 & sdM7 & $ -3.038 \pm 0.004$ & $2839 \pm 6$ & $1.22 \pm 0.05$ & $105.88 \pm 0.34$ & $5.25 \pm 0.05$ & $5-10$ & $38.84 \pm 0.16$\\
  J1444$-$2019 & sdM9 & $-3.457 \pm 0.009$ & $2477 \pm 13$ & $0.99 \pm 0.05$ & $87.35 \pm 0.58$ & $5.35 \pm 0.05$ & $5-10$ & $17.3 \pm 0.17$\\
  J1610-0040\tablenotemark{b} & sdM7 & $-2.97 \pm 0.01$ & $2890 \pm 20$ & $1.27 \pm 0.01$ & $110.72 \pm 0.85$ & $5.23 \pm 0.05$ & $5-10$ & $32.54 \pm 0.36$ \\ 
  J1626+3925 & sdL4 & $-3.787 \pm 0.006$ & $2148 \pm 14$ & $0.9 \pm 0.01$ & $80.2 \pm 1.6$ & $5.39 \pm 0.01$ & $5-10$ & $ 30.78 \pm 0.22$\\
  J2036+5059 & sdM7.5 & $-2.833 \pm 0.011$ & $2983 \pm 22$ & $1.4 \pm 0.01$ & $123.5 \pm 1.3$ & $5.19 \pm 0.05$ & $5-10$ & $43.29 \pm 0.54$\\ 
  \enddata
  \tablecomments{The effective temperature is determined based off age estimates and evolutionary models.}
  \tablenotetext{a}{J1610-0040 is an astrometric binary, thus we caution the reader on the use of these parameters.}
\end{deluxetable*}

\subsection{Fundamental parameters and spectral features of J1256$-$0224 in comparison to the subdwarf sample}

\begin{figure*}
\gridline{\hspace{-0.15cm}\fig{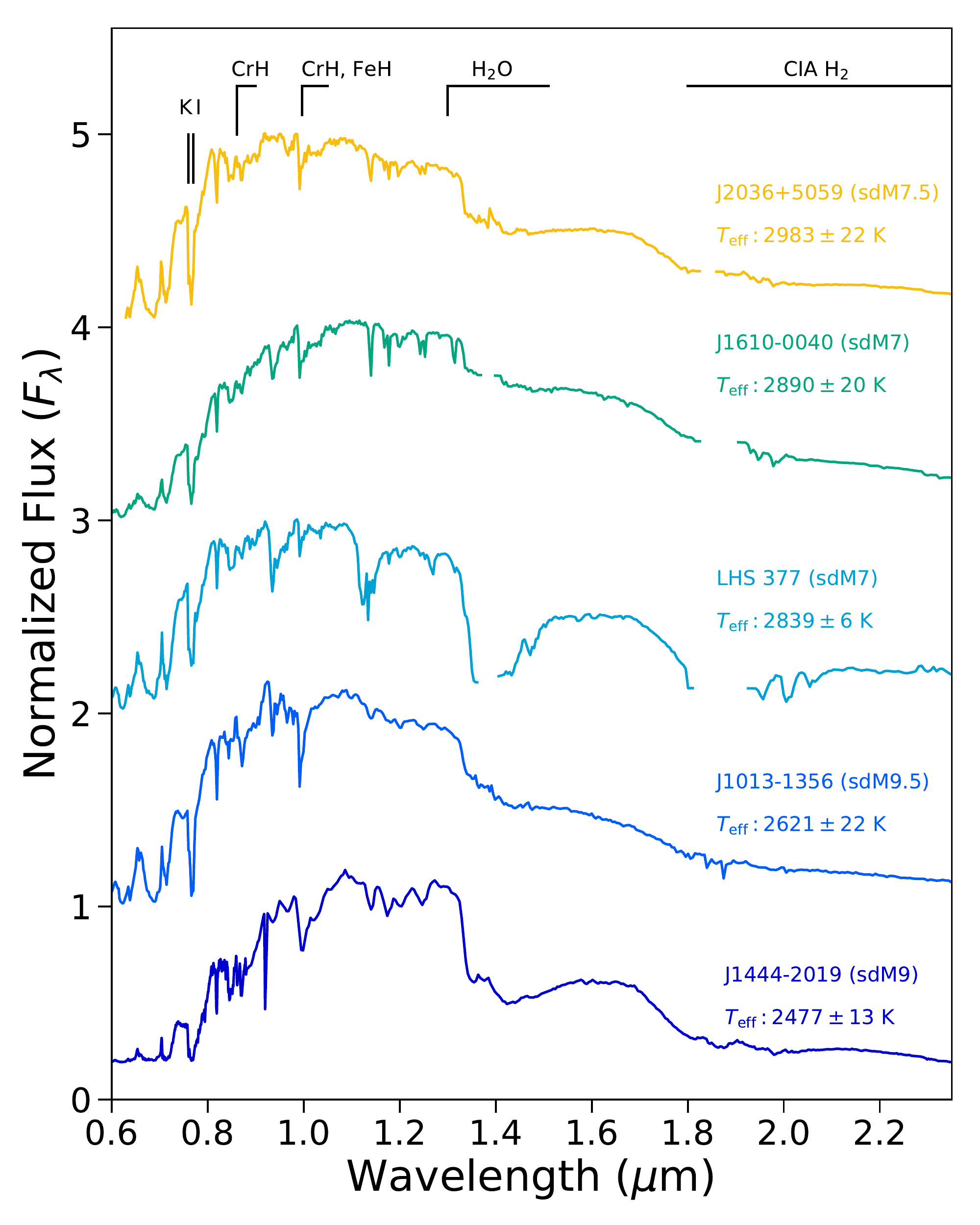}{0.5\textwidth}{\large(a)}
		  \hspace{0.2cm}\fig{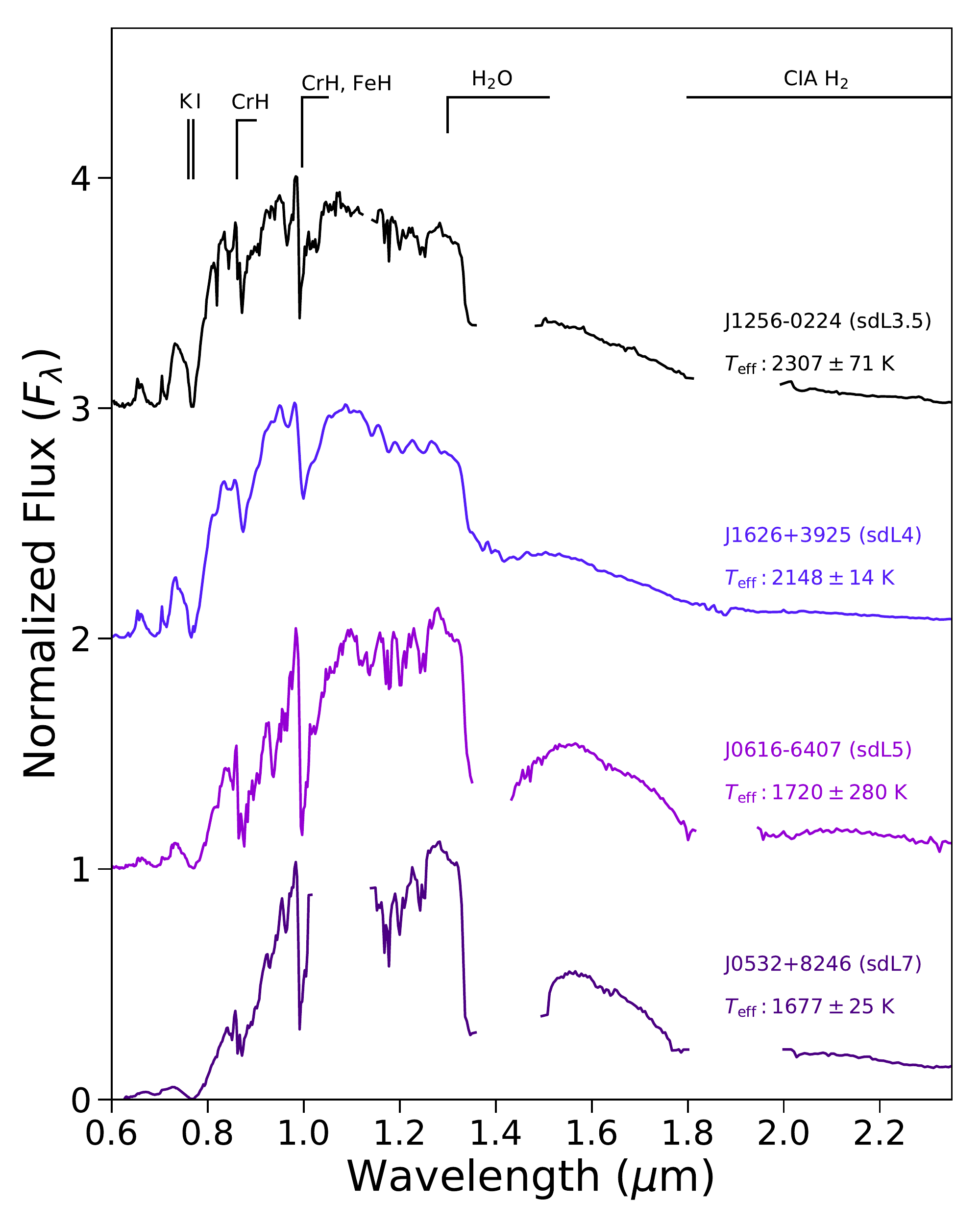}{0.5\textwidth}{\large(b)}}
\caption{Spectra of subdwarfs with parallaxes and spectra over $0.6-2.35\, \upmu$m. Spectra are shifted by a constant and normalized by the average over $0.98-0.988\, \upmu$m. All spectra were resampled to the same dispersion relation using a wavelength-dependent Gaussian convolution. $T_\mathrm{eff}$ decreases from yellow to purple. Note that spectral type does not decrease with $T_\mathrm{eff}$. The optical spectrum of J1425+7102 is not shown in this figure, due to the normalization region. (a) M subdwarfs (b) L subdwarfs}
\label{fig:SubdwarfsFullSpectra}
\vspace{0.5cm} 
\end{figure*}

\begin{figure*}
\centering
 \includegraphics[scale=.845]{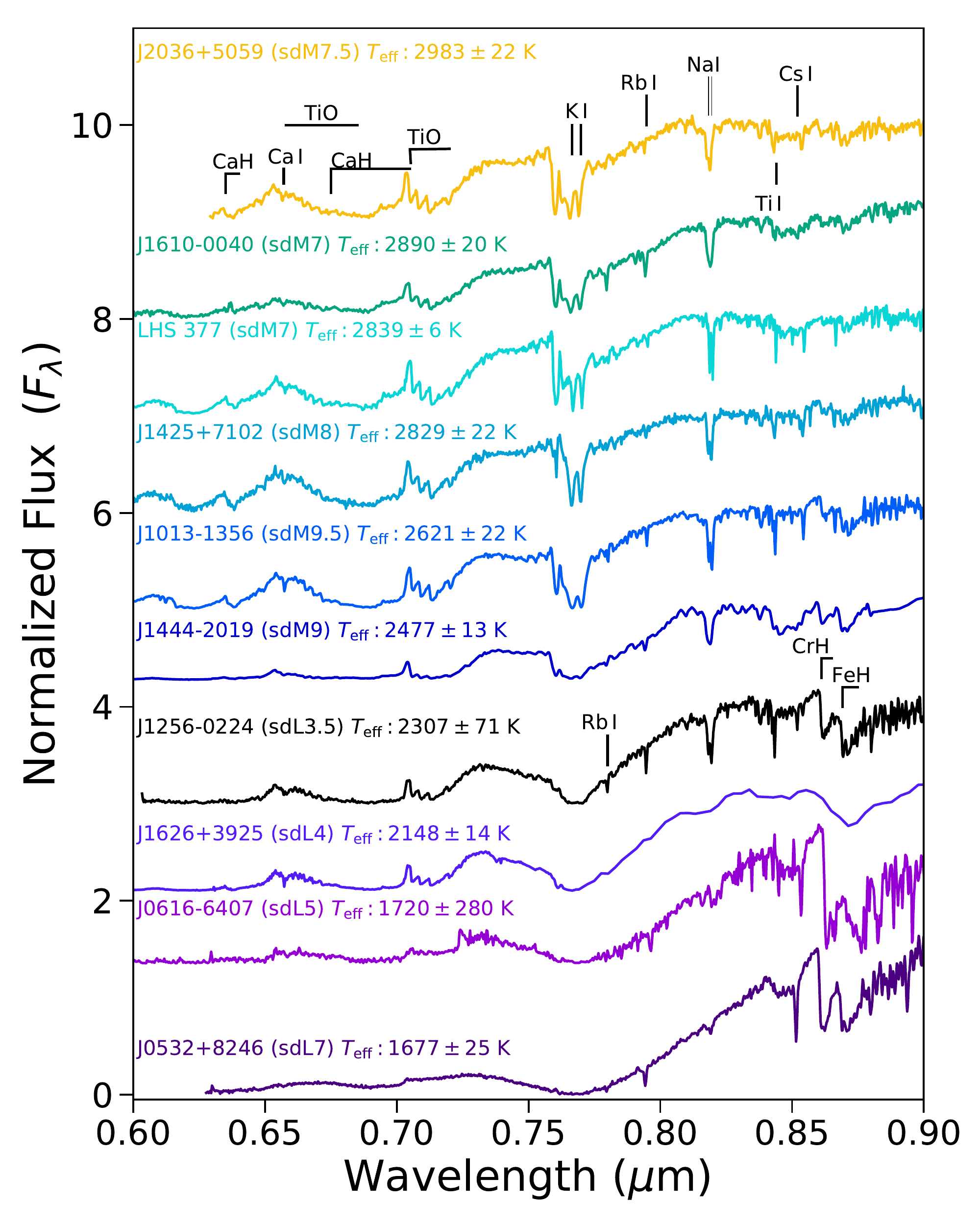}
\caption{Red optical SEDs of subdwarfs with parallaxes and spectra normalized by the average over $0.825-0.840\, \upmu$m. All spectra, except J1444$-$2019 and 1626+3925 which are shown at a lower resolution, were resampled to the same dispersion relation using a wavelength-dependent Gaussian convolution. $T_\mathrm{eff}$ decreases from yellow to purple.}
\label{fig:Subdwarfsredoptical}
\end{figure*}

The semi-empirical fundamental parameters for the subdwarf parallax sample are shown in Table \ref{tab:subdwarfSEDout}.  Figure \ref{fig:SubdwarfsFullSpectra} shows the subdwarf sample displayed in the effective temperature sequence from $0.6-2.2\, \upmu$m with $T_\mathrm{eff}$ decreasing from yellow (hot) to purple (cool), with J1256$-$0224 highlighted in black. Figure \ref{fig:Subdwarfsredoptical} shows a zoom in on the red optical wavelength region with molecular and line features labeled. 

As seen in Figure \ref{fig:SubdwarfsFullSpectra}, our derived subdwarf temperature sequence does not closely follow the spectral type sequence. For the M subdwarfs in particular, subtypes half a type cooler (i.e sd M7.5) are typically found to be warmer than that of the integer type (i.e sd M7). J1013$-$135 sticks out as an sdM9.5 between a hotter sdM7 and a cooler sdM9. This mixed spectral type order indicates that ordering subdwarfs via spectral type is not a good temperature proxy. There are variations of the spectral types for three subdwarfs due to different spectral typing schemes. For example, J1444$-$2019 is typed as an sdM9 in \cite{Scho04c}, in \cite{Kirk16} it is spectral typed as an sdL0, and in \cite{Zhang2017a} as an esdL1. Spectral schemes such as that of \cite{Zhang2017a} attempt to improve upon the relation between spectral type and temperature however, as shown in this work spectral type remains a poor temperature proxy for subdwarfs.

\subsubsection{Variations in subdwarf spectral lines as a function of temperature}\label{Subdwarfspecfeatures}
In Figure~\ref{fig:Subdwarfsredoptical} features are labeled with the objects plotted in decreasing temperature sequence, allowing us to examine how spectral features change as a function of temperature instead of spectral type. As the temperature decreases, the $0.63\, \upmu$m CaH molecular band strengthens slightly from $\sim 3000$~K to 2600~K, then disappears around 2400~K. The \ion{Ca}{1} feature near $0.66\, \upmu$m also strengthens and then becomes less visible as the temperature decreases. The $0.66-0.71\, \upmu$m TiO- and CaH-sculpted region widens and flattens as the temperature decreases, eventually becoming undetectable for J0532+8246. \cite{Kirk14} noted that M/L transition subdwarfs have a noticeable peak between the CaH and TiO bands around $0.70\, \upmu$m, and that the slope of the plateau between approximately $0.73\, \upmu$m and $0.76\, \upmu$m changes from red to blue from types sdM9 to sdL1. The peak-shaped feature between CaH and TiO at $0.71\, \upmu$m is visible for all subdwarfs in our sample. The feature shrinks as the TiO on the right weakens and disappears, leaving only a small bump visible between the CaH and TiO when the temperature decreases to that of J0532+8246. The plateau between $0.73-0.76\, \upmu$m, bounded by TiO on the left and a \ion{K}{1} doublet on the right, is visible down to the temperature of J1626+3925, and is no longer visible below that temperature due to the broadening of the \ion{K}{1} doublet. The \ion{K}{1} doublet broadens as the atmospheric temperature cools. The individual lines of the \ion{K}{1} doublet are only visible above roughly the temperature of J1013$-$1356, $\sim 2600$~K. The $0.78\, \upmu$m \ion{Rb}{1} line is clearly visible for J1256$-$0224 only. The $0.79\, \upmu$m \ion{Rb}{1} line however, strengthens as the temperature decreases to that of J1256$-$0224, but then appears to decrease in strength for J0616$-$6407 and J0532+8246. The \ion{Na}{1} doublet is of similar depth for all objects between $\sim 3000-2300$~K, and isn't easily visible below those temperatures. The \ion{Ti}{1} line near $0.84\,\upmu$m strengths as the temperature cools down to that of J1256$-$0224 and then appears to weaken below $\sim2300$~K. The \ion{Cs}{1} line at $0.8521\,\upmu$m however, appears to strengthen as the temperature cools. Both the CrH and FeH molecular features near $0.87\, \upmu$m deepen as the temperature decreases. For subdwarfs, all of these optical spectral features appear to be primarily driven by temperature.

\subsubsection{Analysis of the $J$-band \ion{K}{1} doublets in the subdwarf sample }\label{SubdwarfKIlines}
\begin{figure}
\hspace{-0.25cm}
 \includegraphics[scale=.43]{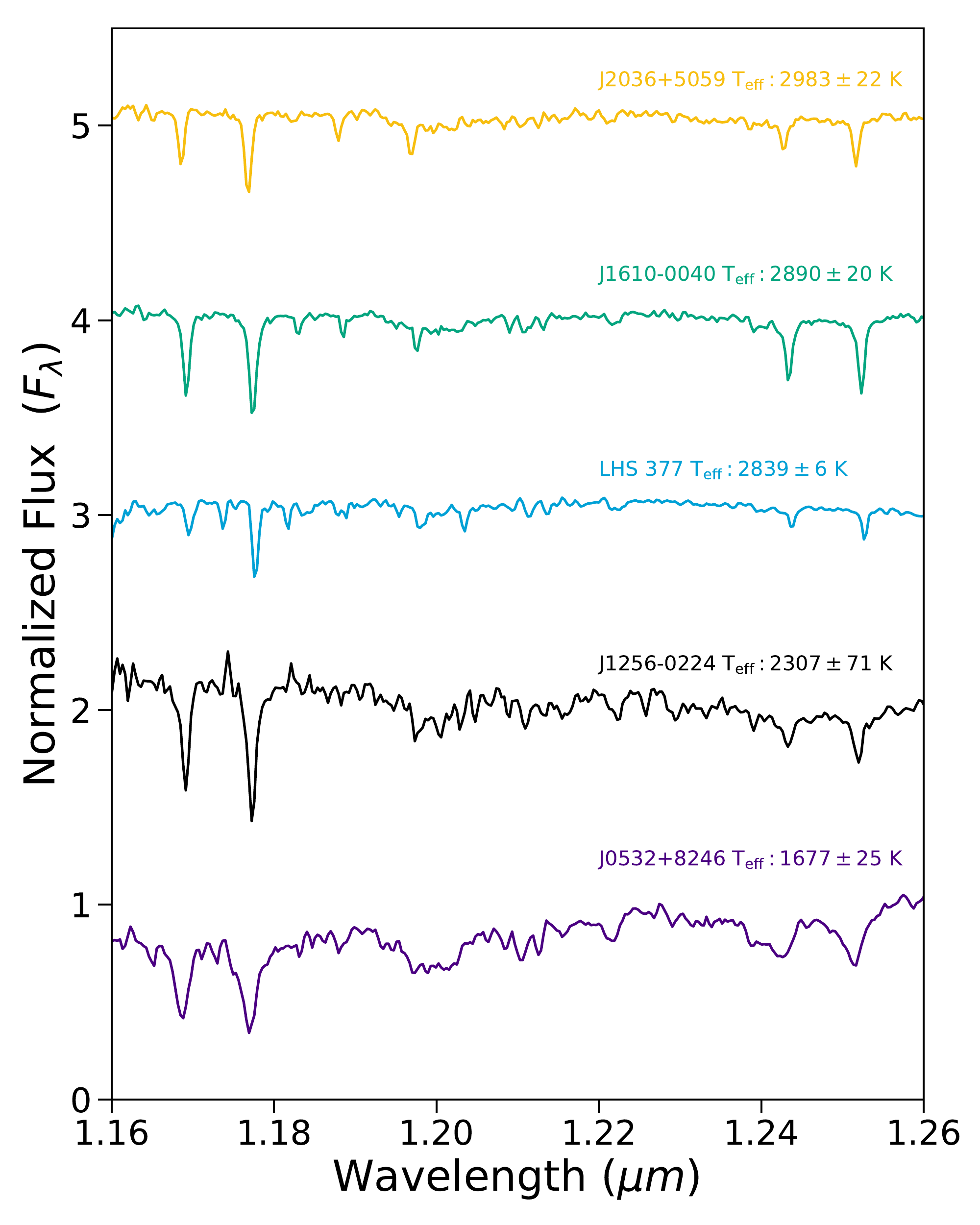} 
\caption{$J$ band between $1.16-1.26\, \upmu$m focused on the \ion{K}{1} doublets of the subdwarfs with medium-resolution spectra. All spectra were resampled to the same dispersion relation using a wavelength-dependent Gaussian convolution.}
\label{fig:SubdwarfJ}
\end{figure}

We examined the $J$ band for subdwarfs with medium-resolution NIR data in order to put the \ion{K}{1} doublets of J1256$-$0224 in context with other subdwarfs. The blue \ion{K}{1} doublet, near $1.17\, \upmu$m, of J1256$-$0224 indicates high gravity while the red doublet, near $1.25\, \upmu$m, conversely indicates a low surface gravity. Figure \ref{fig:SubdwarfJ} shows the 5 subdwarfs that have medium-resolution $J$ band data across the $1.16-1.26\, \upmu$m region. In general, the blue \ion{K}{1} doublet is deeper than the red doublet for all of these subdwarfs. The depths of both \ion{K}{1} doublets do not change monotonically as a function of temperature. The doublets of J1256$-$0224 show the largest difference in depth between the two sets, compared to the other subdwarfs, and its blue doublet is the deepest overall. LHS 377 is the only subdwarf in this set to have a large difference in depth between the first and second line in the $1.17\, \upmu$m \ion{K}{1} doublet. Obtaining more medium-resolution $J$ band spectra of subdwarfs would help determine the underlying cause of these conflicting indications of gravity with \ion{K}{1} doublets. Thus, we conclude that the $1.25\, \upmu$m \ion{K}{1} doublet is a poor indicator of gravity for subdwarfs, while the $1.17\, \upmu$m doublet needs to be studied more to determine if it is still a suitable gravity indicator for subdwarfs. 

\subsection{Subdwarf fundamental parameters compared to field age and young dwarfs}
\begin{figure*}
\centering
 \includegraphics[scale=.57]{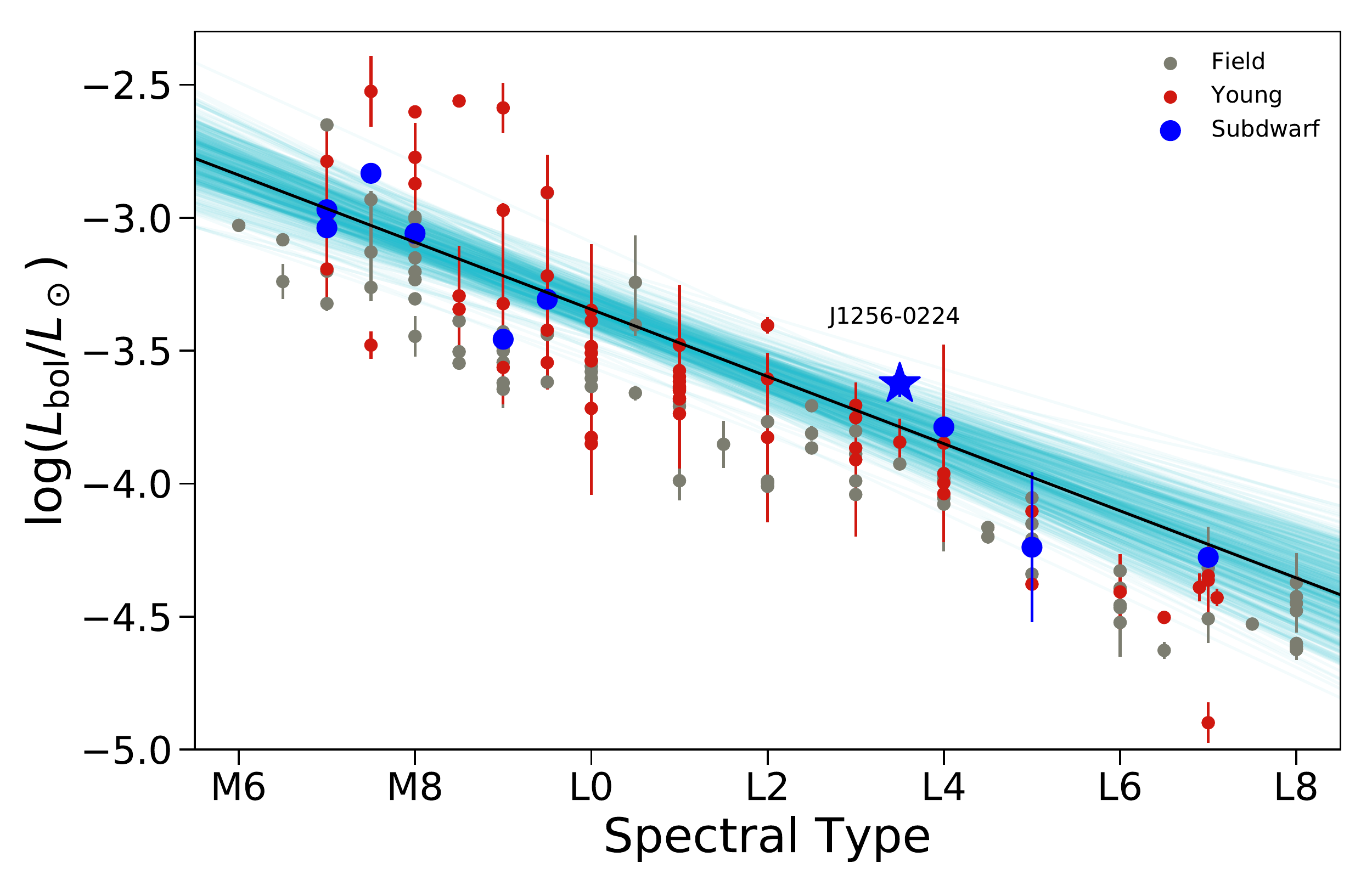} 
\caption{Spectral Type vs $L_\mathrm{bol}$ for subdwarfs (blue), field objects (grey), and low gravity objects (red). The field objects come from \cite{Fili15} and the low gravity objects are from \cite{Fahe16}. $L_\mathrm{bol}$ was calculated for the field- and low-gravity objects using the same method as used for the subdwarfs. A linear best fit line is shown in black, with 1000 random samples from the MCMC shown in aqua.}
\label{fig:SptvLbol}
\end{figure*}

\begin{figure*}
\centering
 \includegraphics[scale=.57]{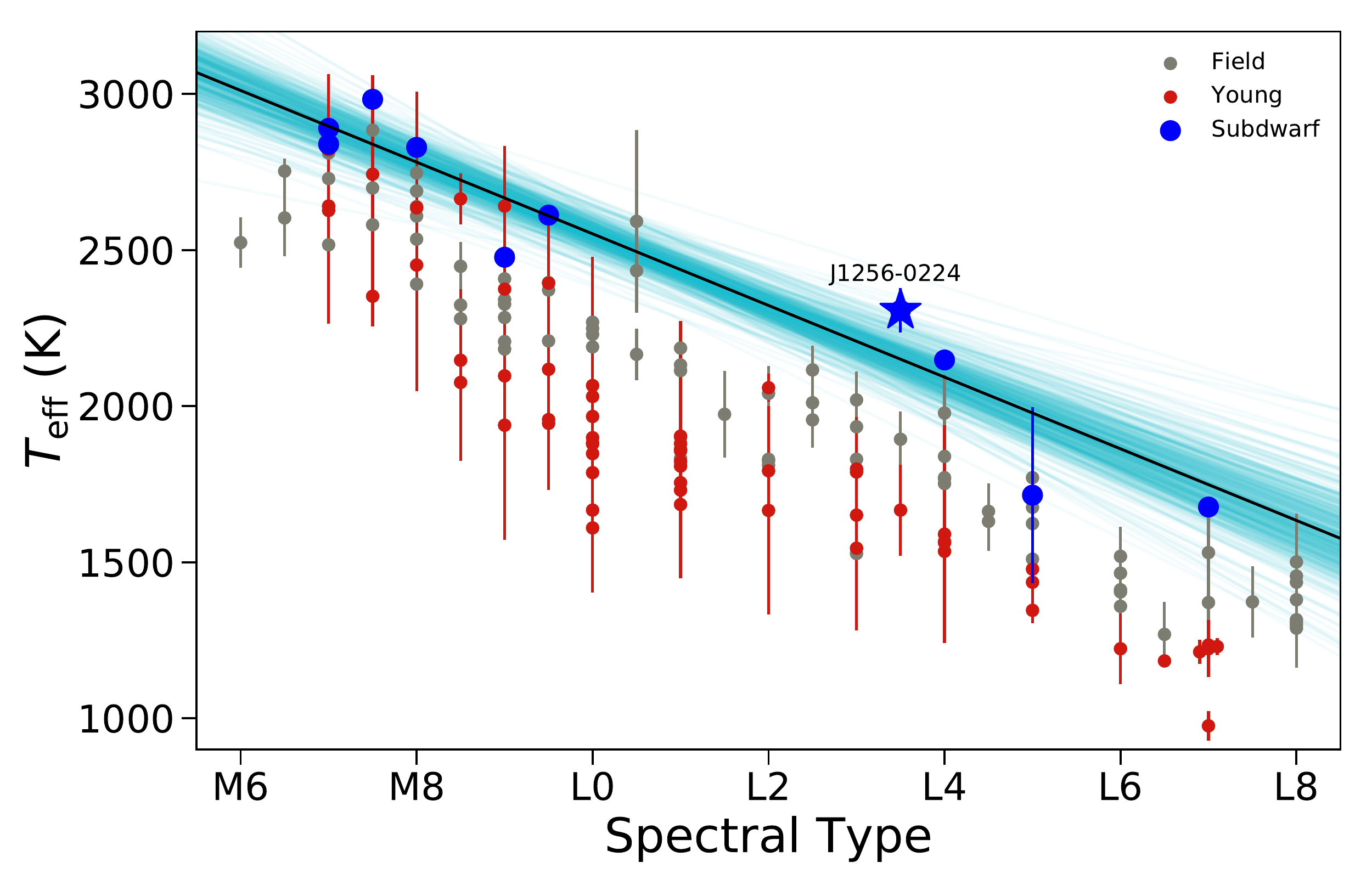}
\caption{Spectral Type vs $T_\mathrm{eff}$ for subdwarfs (blue), field objects (grey), and low gravity objects (red). The field objects come from \cite{Fili15} and the low gravity objects are from \cite{Fahe16}. Effective temperatures for field- and low-gravity objects were calculated using the same method as done for the subdwarfs. A linear best fit line is shown in black, with 1000 random samples from the MCMC shown in aqua.}
\label{fig:SptvTeff}
\end{figure*}

\begin{figure*}[!htb]
\centering
 \includegraphics[scale=.57]{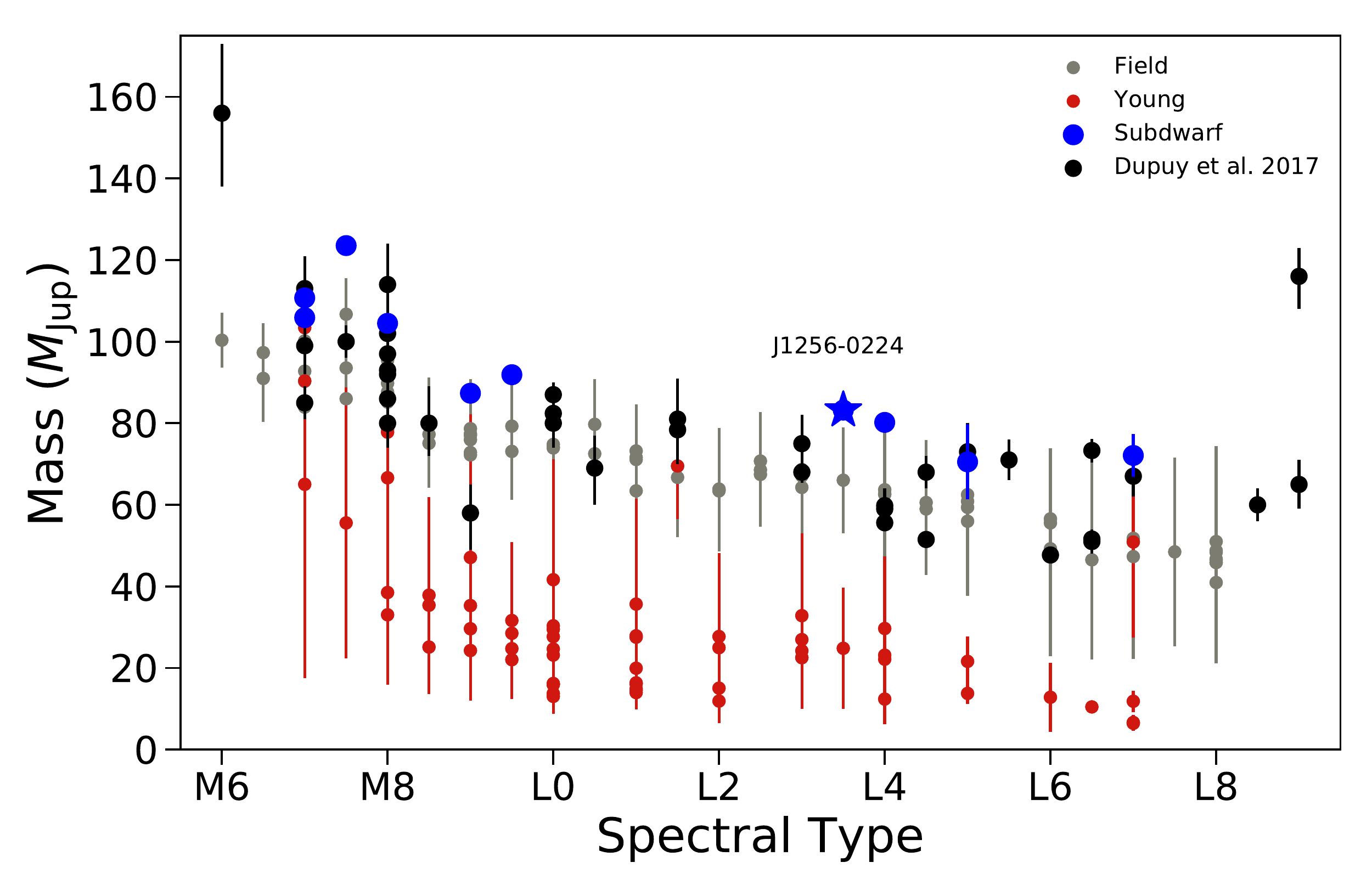} 
\caption{Spectral Type vs Mass for subdwarfs (blue), field objects (grey), low gravity objects (red), and sources from the \cite{Dupu17} sample in black. Mass was calculated for the field and low gravity objects using the same method as the subdwarfs, where the mass is determined based off of the estimated model radius. The \cite{Dupu17} sample is of empirical mass measurements. The IR spectral type was used for objects from \cite{Fahe16} if no optical spectral type was available.}
\label{fig:SptvMass}
\end{figure*}

Figure \ref{fig:SptvLbol} compares spectral type versus $L_\mathrm{bol}$ for subdwarfs, field- and low-gravity objects. The subdwarfs are entangled with the field- and low-gravity sequences. The unclear bolometric luminosity separation at the same spectral type for these populations indicates that spectral typing is not a good proxy for the effective temperature of subdwarfs. This is also the case for low-gravity objects \citep{Fili15, Fahe16}. A linear fit for the subdwarfs (see Figure~\ref{fig:SptvLbol}, and Table~\ref{tab:FitCoeffs} for fit coefficients) appears to be located slightly above the field sequence, but since we only have 11 subdwarfs in this sample, it is difficult to determine if this fit is the best assessment for the entire subdwarf population. 

As seen in Figure \ref{fig:SptvTeff}, the spectral type versus $T_\mathrm{eff}$ of the subdwarfs compared to field- and low-gravity sources show the subdwarfs above the field sequence, with warmer temperatures than field objects of the same spectral type. The low-gravity sources lie below the field sequence, hence are cooler (e.g. see \citealt{Fili15}; \citealt{Fahe16}). The difference in effective temperature for low-gravity sources appears to become more differentiated at later spectral types. The subdwarfs may also follow this trend, however more subdwarfs of type sdL3 and later are needed to verify this trend. A linear fit was determined for the subdwarf population, which shows subdwarfs lying above the field sequence. The fit can be seen in Figure~\ref{fig:SptvTeff} and the fit coefficients are listed in Table~\ref{tab:FitCoeffs}. 

An interesting aspect of Table \ref{tab:subdwarfSEDout} is that the subdwarf sample contains four brown dwarfs, while the other 6 are stars. Assuming ages $> 5$~Gyr,and using [Fe/H] values and equation 1 from \cite{Zhang2017a} for LHS 377, J1013$-$1356, J1256$-$0224, J1626+3925, J0616$-$6407, and J0532+8246, the substellar transition occurs at temperatures between $\sim 2280-2510$~K, for metallicites between $-1.2$ and $-1.8$~dex based on the hydrogen burning minimum mass for these metallicites. The evolutionary model radii of the subdwarf sample is predicted to range from $0.84-1.4\, R_\mathrm{Jup}$, with radii below $\sim0.94\, R_\mathrm{Jup}$ corresponding to the brown dwarfs in the sample. 

Figure \ref{fig:SptvMass} shows spectral type versus semi-empirically derived masses for subdwarfs, field- and low-gravity objects. Under close inspection, some subdwarfs of different spectral types are found to have similar masses. J1444$-$2019 and J1256$-$0224 are extreme examples which differ by 5 spectral types, but have similar masses. The masses we display are determined based on assumed age and thus radius. For comparison, the dynamical masses from \cite{Dupu17} are plotted in black, which does not include subdwarfs. The \cite{Dupu17} sample fits well to the semi-empirically derived field sequence, with only a few outliers far above the field. This indicates that our model mass estimations have a good relative accuracy for the field. The subdwarfs do not lie very far above the field sequence or dynamical mass measurements, with a few of the \cite{Dupu17} objects in the same positions as the subdwarfs. Thus dynamical mass measurements of subdwarfs are needed to provide better information on how subdwarfs distinguish themselves from the field sequence, when comparing spectral type and mass. At this time there is only one close subdwarf binary, J1610$-$0040 an sdM7, whose dynamical mass measurements are estimated in \cite{Kore16} as $0.09-0.10\, M_\odot$ for the primary and $0.06-0.075\, M_\odot$ for the secondary. Our estimated total mass of the J1610$-$0040 system is less than the value determined from \cite{Kore16} by an appreciable amount indicating that low metallicity models may be underestimating masses. The components of J1610$-$0040 can be placed on Figure~\ref{fig:SptvMass} for comparison to \cite{Dupu17} once spectral types are determined.

\subsection{Subdwarf absolute magnitudes comparisons}

\begin{figure}
\gridline{\hspace{-0.1cm}\fig{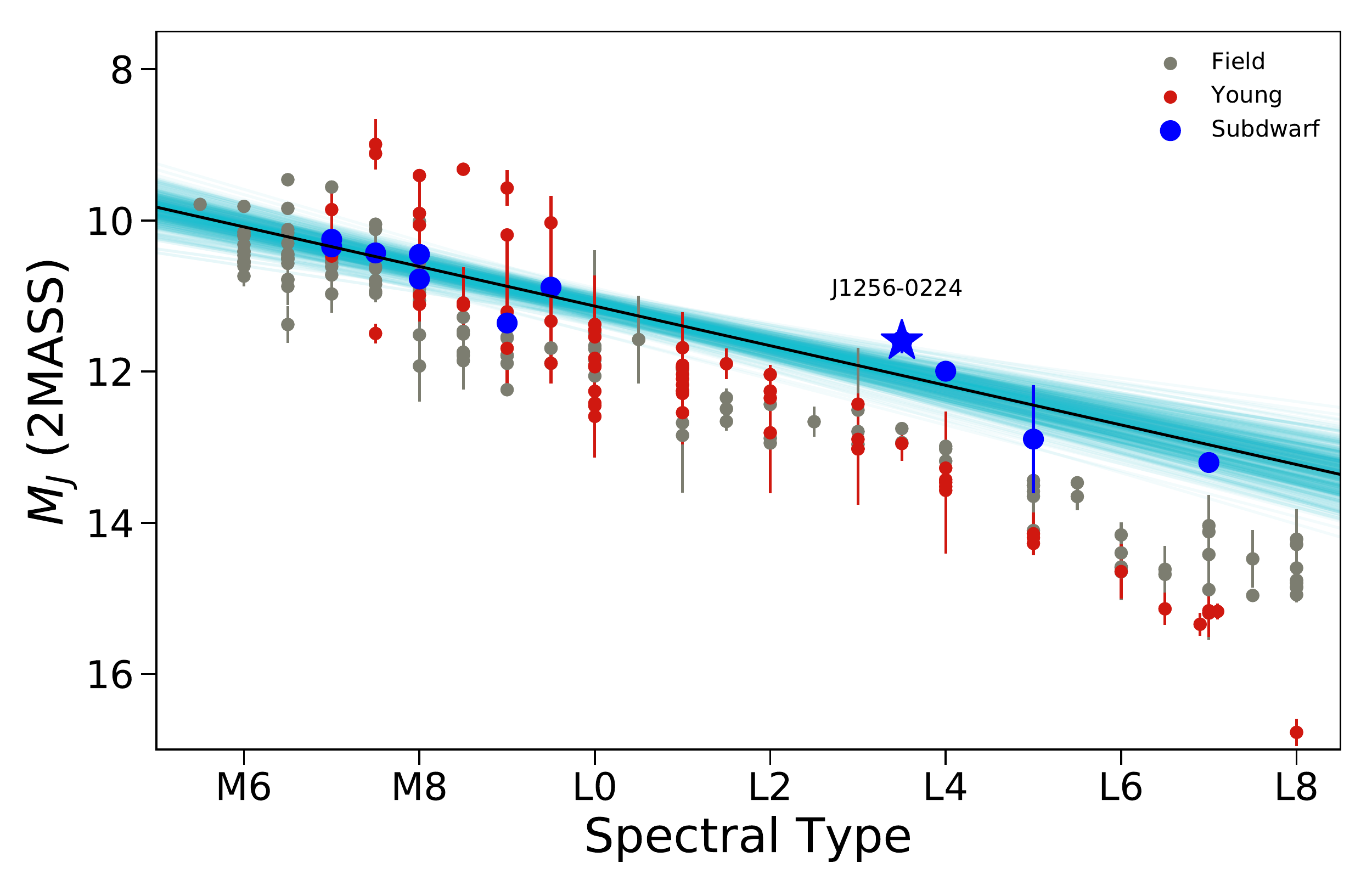}{0.48\textwidth}{\large(a)}}
\gridline{\hspace{-0.1cm}\fig{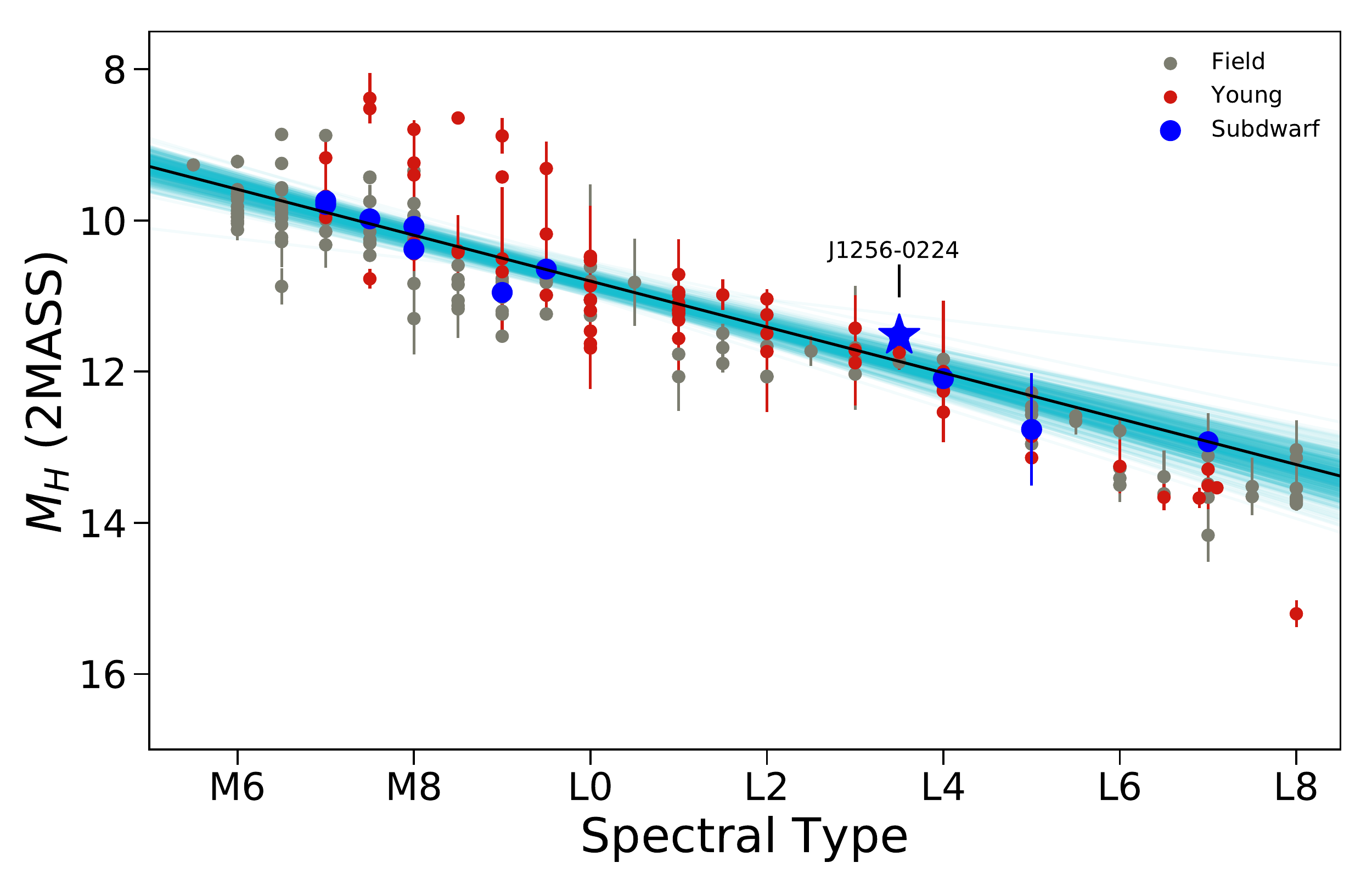}{0.48\textwidth}{\large(b)}}
\gridline{\hspace{-0.1cm}\fig{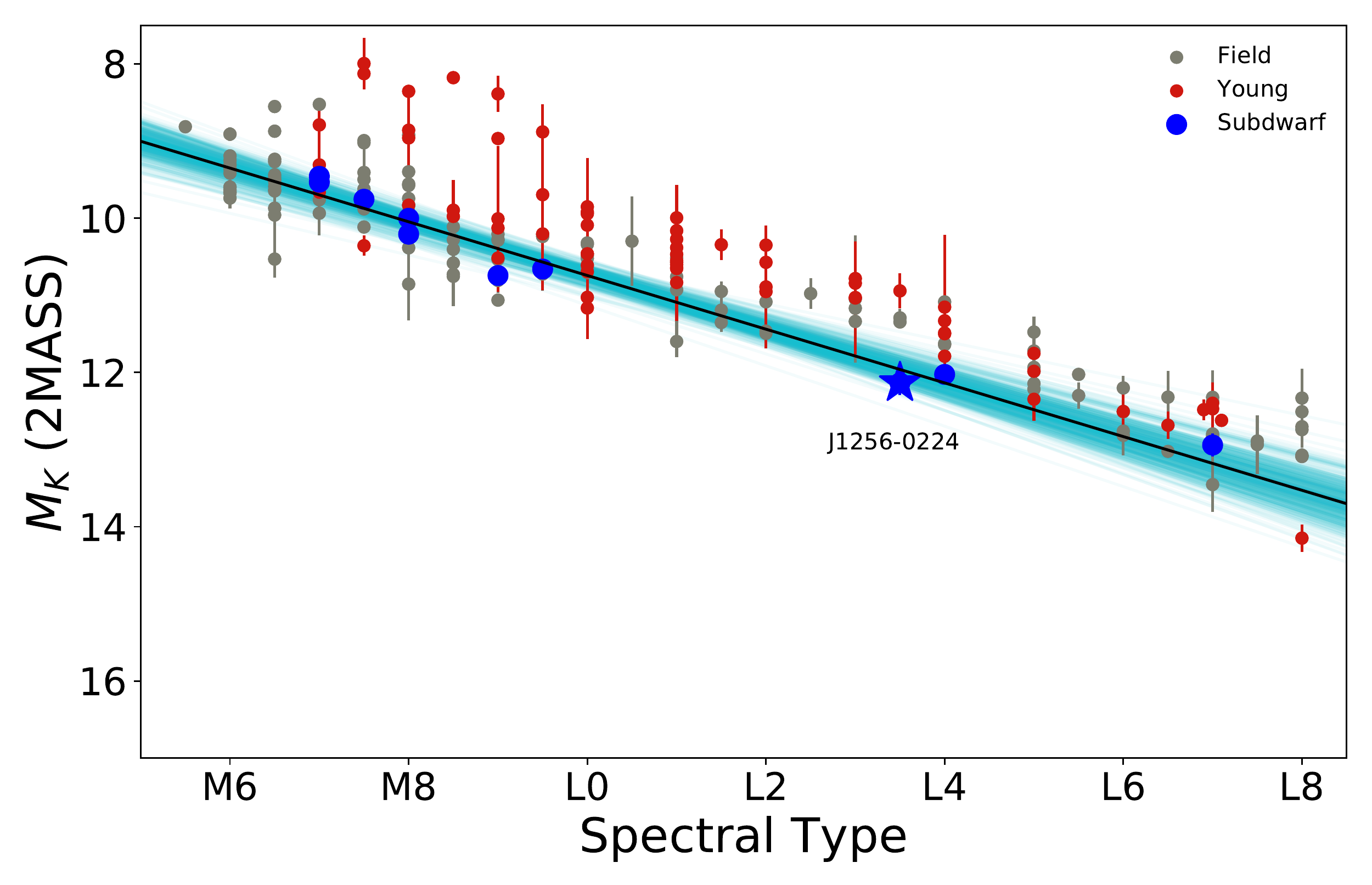}{0.48\textwidth}{\large(c)}}
\caption{Spectral type versus 2MASS absolute magnitudes. Field objects are shown in grey, low gravity in red and subdwarfs in blue. A linear best fit line is shown in black, with 1000 random samples from the MCMC shown in aqua. (a) Spectral Type vs $M_J$\, (b) Spectral Type vs $M_H$ \, (c) Spectral Type vs $M_{Ks}$}
\label{fig:AbsMagsJHK}
\end{figure}

\begin{figure}
\gridline{\hspace{-0.1cm}\fig{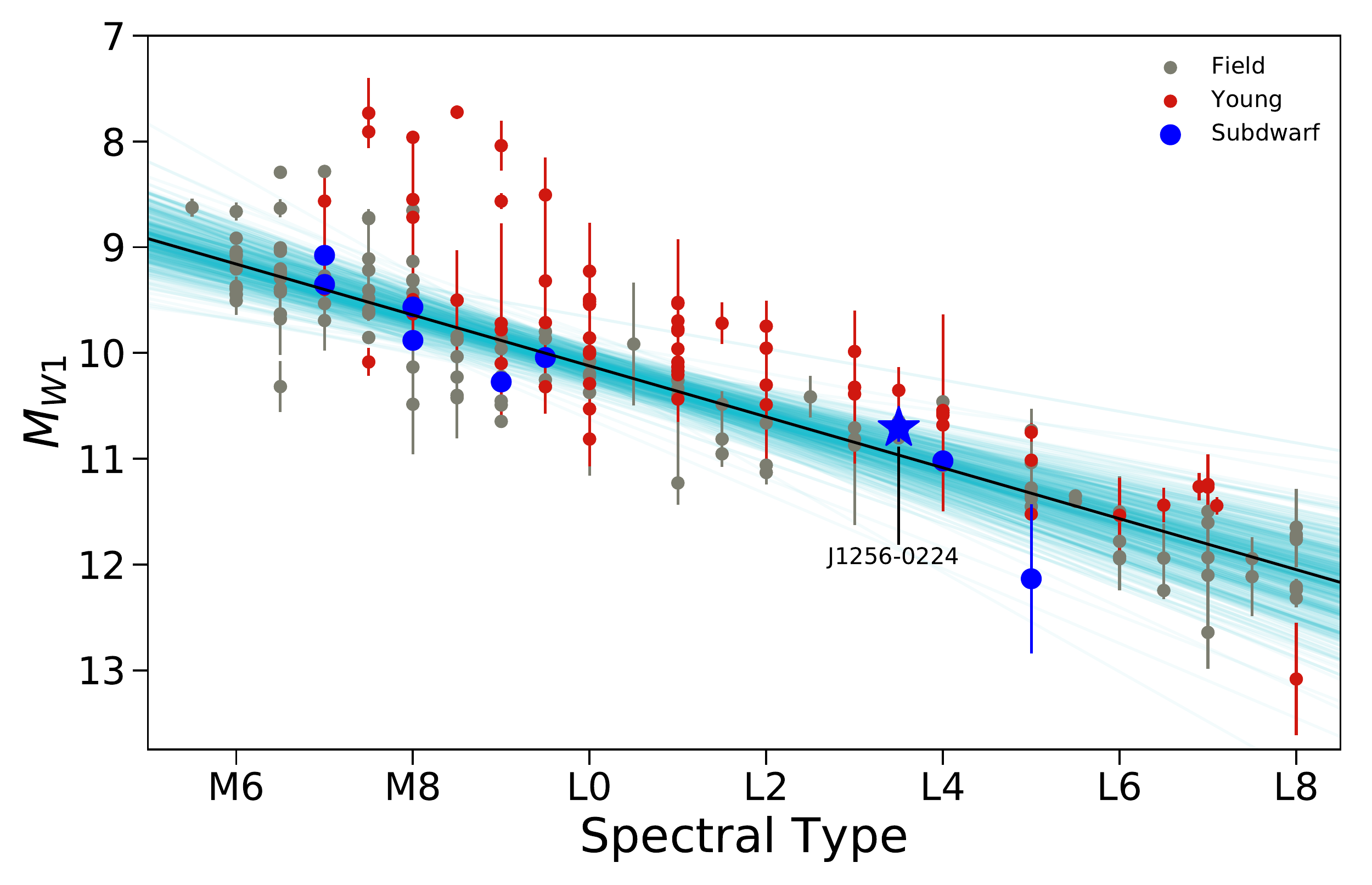}{0.48\textwidth}{\large(a)}}
\gridline{\hspace{-0.1cm}\fig{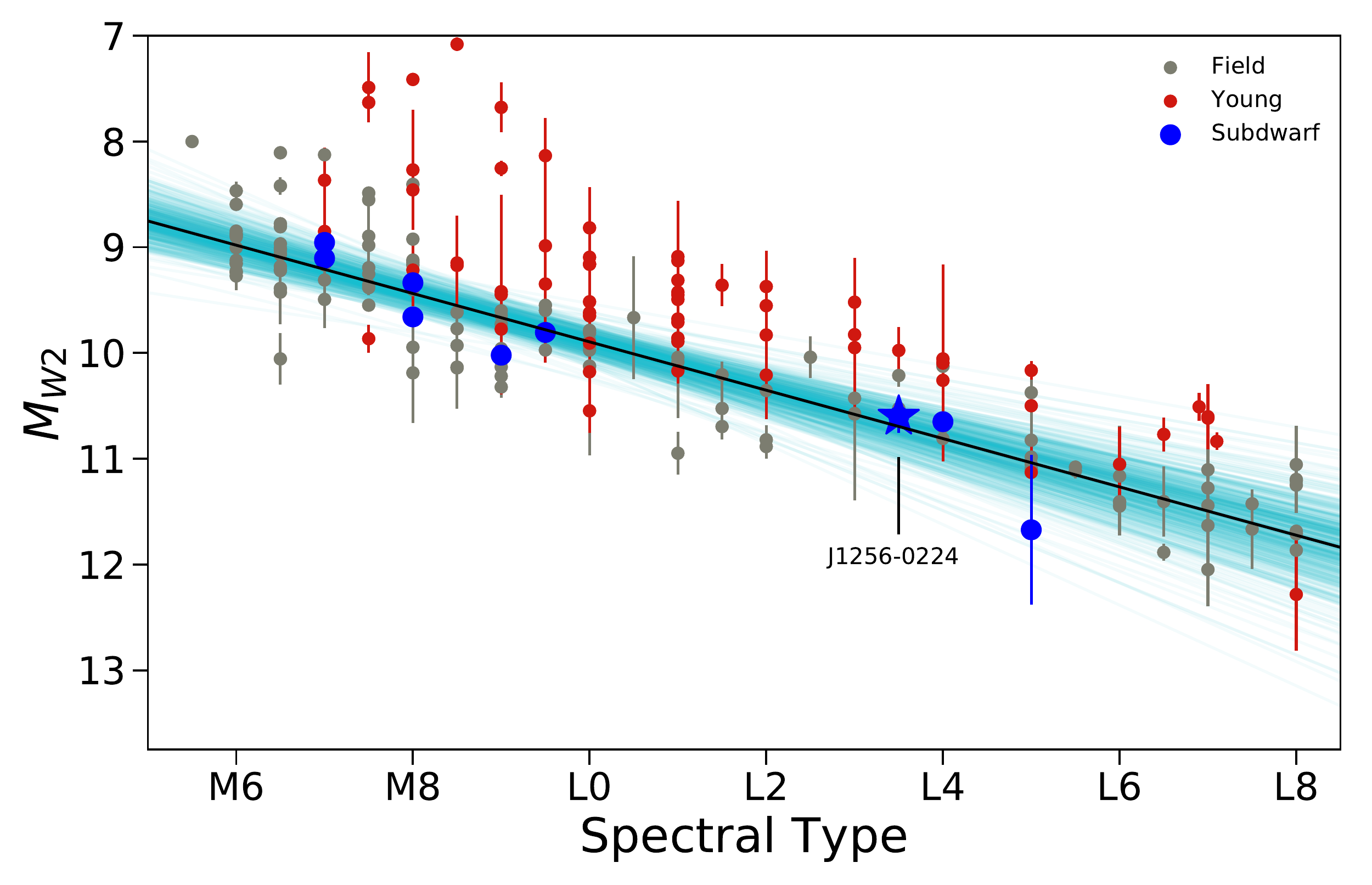}{0.48\textwidth}{\large(b)}}
\caption{Spectral type versus WISE absolute magnitudes. Field objects in grey, low gravity in red and subdwarfs in blue. A linear best fit line is shown in black, with 1000 random samples from the MCMC shown in aqua. (a) Spectral Type vs $M_{W1}$ \, (b) Spectral Type vs $M_{W2}$}
\label{fig:AbsMagsW1W2}
\end{figure}

As an extension of the spectral type to absolute magnitude relations for subdwarfs in Figure 13 of \cite{Fahe12}, Figures \ref{fig:AbsMagsJHK} and \ref{fig:AbsMagsW1W2} display spectral type versus absolute magnitude in 2MASS $J$, $H$, $K$, and WISE $W1$ and $W2$ photometric bands. Each absolute magnitude band versus spectral type was fit with a polynomial for the subdwarf sample. The coefficients of these fits are listed in Table \ref{tab:FitCoeffs}. As mentioned in \cite{Burg08d} and \cite{Fahe12}, the subdwarfs move from being brighter in $M_J$ to normal or slightly fainter than the field at $M_{K}$, with this effect less prominent for the late-M subdwarfs. From $M_{K}$ to $M_{W2}$ we see the subdwarfs remain normal, with the exception of the L subdwarf J0616$-$6407 which becomes slightly fainter than the field. From $M_{K}$ to $M_{W2}$ J1256$-$0224 however, remains normal compared to the field.

\begin{deluxetable}{l c c c}
\tablecaption{Polynomial relations for Subdwarfs \label{tab:FitCoeffs}}
\tablehead{\colhead{Relation} & \colhead{C$_0$\tablenotemark{a}} & \colhead{C$_1$} & \colhead{$\sigma$\tablenotemark{b}}} 
  \startdata
  $M_J$ & $0.263 \pm 0.027$ & $8.49 \pm 0.28$ & $0.258 \pm 0.075$\\
  $M_H$ & $0.304 \pm 0.026$ & $7.77 \pm 0.27$ & $0.236 \pm 0.082$\\ 
  $M_K$ & $0.344 \pm 0.024$ & $7.29 \pm  0.25$ & $0.208 \pm 0.074$\\             
  $M_{W1}$ & $0.241 \pm 0.043$ & $7.72 \pm 0.42$ & $0.274 \pm 0.099$\\  
  $M_{W2}$ & $0.228 \pm 0.039$ & $7.62 \pm 0.38$ & $0.238 \pm 0.088$\\          
  $T_\mathrm{eff}$ &$-117 \pm 13$ & $3721 \pm 132$ & $108 \pm 39$\\ 
  $L_\mathrm{bol}$ &$-0.125 \pm 0.015$ & $-2.10 \pm 0.16$ & $0.144 \pm 0.048$ \\
  \enddata
\tablecomments{All coefficients and intrinsic dispersions were determined using a Markov chain Monte Carlo (MCMC) calculation with a Gaussian prior, 3 walkers, and 1000 iterations.}
\tablenotetext{a}{$y = C_0 x + C_1$}
\tablenotetext{b}{Intrinsic dispersion}
\end{deluxetable}

\section{Conclusions}
In this work we present the distance-calibrated SED of J1256$-$0224 and compared it to objects of the same effective temperature and bolometric luminosity. Using derived fundamental parameters, we show that the best comparison objects are not of the same spectral type (L3.5), but instead are field- and low-gravity objects 3 to 4 subtypes earlier (M9-L0). We expect a cloudless atmosphere for J1256$-$0224, based on the larger amount of flux visible in the optical and the reduced flux in the $H$, $K$, and mid-infrared bands. Comparing spectral features of the effective temperature sample in the near-infrared subdwarfs have stronger FeH and \ion{K}{1} features, and visible lines of \ion{Ti}{1}. The \ion{K}{1} doublets in the $J$ band for J1256$-$0224 show indications of low gravity for the $1.17\,\upmu$m doublet and high gravity for the $1.25\,\upmu$m doublet. We note that the subdwarfs in our comparison sample with medium-resolution $J$-band data also show both low and high gravity for the \ion{K}{1} doublets and thus we do not believe that the $1.25\,\upmu$m doublet is a good indicator of surface gravity for subdwarfs. We also see an indication of CO in the $K$ band for J1256$-$0224, which previously went undetected in the SpeX spectrum from \cite{Burg09a}. 

In order to place J1256$-$0224 in context with subdwarfs, we present distance-calibrated SEDs of the 11 subdwarfs typed sdM7 and later with parallaxes. The SEDs are displayed in a decreasing temperature sequence in order to understand how subdwarf spectral lines change with temperature. As the temperature cools, the $0.73-0.76$\,\AA\ plateau feature moves from a red slope to a blue slope, the \ion{K}{1} doublet near $0.77\,\upmu$m broadens, and CrH and FeH strengthen in the red optical. The spectra of J1013$-$1356 and J1256$-$0224 have \ion{Ca}{1} absorption at $6571$\,\AA\ visible in the red optical. We show that spectral type is not a good proxy for subdwarf effective temperature and note in our sample we see half-subtypes are warmer than integer subtypes when calculating $T_\mathrm{eff}$ via SED fitting. Sequences were determined for subdwarfs for spectral type versus $L_\mathrm{bol}$, spectral type versus $T_\mathrm{eff}$, spectral type versus mass and are compared to the field- and low-gravity sequences for each. We show that all populations are indistinguishable based on only spectral type versus $L_\mathrm{bol}$, however populations separate using spectral type versus $T_\mathrm{eff}$, with subdwarfs $\sim300$~K warmer on average than equivalent spectral typed field objects. We also expanded the spectral type versus absolute magnitude sequences down to $W1$ and $W2$, where subdwarfs are normal to slightly fainter than the field dwarfs. Subdwarfs are an important link in understanding how brown dwarf atmospheres have changed over time and thus how metallicity affects spectral features.

\acknowledgments
The authors would like to thank Adam Burgasser, Arvind Rajpurohit, Michael Cushing, Ralf-Dieter Scholz, and S\'ebastien L\'epine for the spectra they provided to improve the coverage of our subdwarf SEDs. We also thank ZengHua Zhang for his useful comments on the manuscript and NIR spectrum of J0616$-$6407, as well as our anonymous referee for the helpful manuscript comments. We thank the Magellan telescope operators for their help in collecting FIRE spectra. EG thanks David Rodriguez and Joe Filippazzo for their help with understanding and modifying SEDkit for this work. This research was partially supported by the RISE Program at Hunter College, grant \#GM060665, and by the NSF under Grant No. AST-1614527 and Grant No. AST-1313278. EG thanks the LSSTC Data Science Fellowship Program, her time as a Fellow has benefited this work. This research has made use of the BDNYC Data Archive, an open access repository of M, L, T and Y dwarf astrometry, photometry and spectra. This paper includes data gathered with the 6.5 meter Magellan Telescopes located at Las Campanas Observatory, Chile. This publication makes use of data products from the Two Micron All Sky Survey, which is a joint project of the University of Massachusetts and the Infrared Processing and Analysis Center/California Institute of Technology, funded by the National Aeronautics and Space Administration and the National Science Foundation. This publication makes use of data products from the Wide-field Infrared Survey Explorer, which is a joint project of the University of California, Los Angeles, and the Jet Propulsion Laboratory/California Institute of Technology, funded by the National Aeronautics and Space Administration. This work has made use of data from the European Space Agency (ESA) mission {\it Gaia} (\url{https://www.cosmos.esa.int/gaia}), processed by the {\it Gaia} Data Processing and Analysis Consortium (DPAC, \url{https://www.cosmos.esa.int/web/gaia/dpac/consortium}). Funding for the DPAC has been provided by national institutions, in particular the institutions participating in the {\it Gaia} Multilateral Agreement.

\facility{Magellan:Baade (FIRE)}
\software{SEDkit (\url{https://github.com/hover2pi/SEDkit}), emcee \citep{emcee}}

\bibliographystyle{yahapj}
\bibliography{references}

\end{document}